\documentclass[11pt]{article}
\usepackage{latexsym,cite,amssymb}
\makeatletter

\@addtoreset{equation}{section}
\makeatother

\input amssym.def
\input amssym.tex

\newcommand{\bh}{{\bar{h}}}

\newcommand{\bvarphi}{\bar{\varphi}}

\newcommand{\halpha}{{\hat{\alpha}}}
\newcommand{\hbeta}{{\hat{\beta}}}
\newcommand{\hgamma}{{\hat{\gamma}}}
\newcommand{\hdelta}{{\hat{\delta}}}

\newcommand{\hchi}{\hat{\chi}}

\newcommand{\he}{\hat{e}}

\newcommand{\rc}{{\rm c}}

\newcommand{\half}{{{\textstyle\frac{1}{2}}}}

\newcommand{\be}{\begin{equation} }
\newcommand{\ee}{\end{equation} }
\newcommand{\ba}{\begin{array}}
\newcommand{\ea}{\end{array}}

\newcommand{\SO}{\mbox{SO}}

\newcommand\cC{{\cal C}}
\newcommand\cD{{\cal D}}

\newcommand\cH{{\cal H}}
\newcommand\cI{{\cal I}}

\newcommand\cL{{\cal L}}

\newcommand\cN{{\cal N}}

\newcommand\cQ{{\cal Q}}

\newcommand\cS{{\cal S}}
\newcommand\cT{{\cal T}}
\newcommand\cV{{\cal V}}
\newcommand\cZ{{\cal Z}}

\newcommand\fD{{\mathfrak{D}}}

\newcommand\fL{{\mathfrak{L}}}

\newcommand\Rs{R}
\newcommand\on{\!\mathrm{on}}

\newcommand\cone{\!\mathrm{cone}}
\newcommand\rmin{\mathrm{min}}
\newcommand\rmax{\mathrm{max}}

\newcommand\vx{v}

\newcommand\tr{{\rm tr}}

\def\I_N{{1_{\scriptscriptstyle N\times N}}}

\begin{document}
\begin{titlepage}
\title{\vskip -60pt
{\small
\begin{flushright}
hep-th/yymmnnn\\
MPP-2006-129
\end{flushright}}
\vskip 20pt Topological twisting of conformal supercharges\\
~}
\author{Jeong-Hyuck Park${}^{\dagger}$ and Dimitrios Tsimpis${}^{\ast}$}
\date{}
\maketitle
\vspace{-1.0cm}
\begin{center}
~~~\\
${}^{\dagger}$Department of Physics, Sogang University, C.P.O. Box
1142, Seoul 100-611, Korea\\
~{}\\
${}^{\ast}$Max-Planck-Institut f\"{u}r Physik, F\"{o}hringer Ring 6,
80805  M\"{u}nchen,  Germany\\
~{}\\
{\small{Electronic correspondence: {{{park@sogang.ac.kr}}},
{{{tsimpis@mppmu.mpg.de}}}}}\\
~~~\\
~~~\\
\end{center}
\begin{abstract}
\noindent Putting a twisted version of $\cN=4$ super Yang-Mills on a
curved four-dimensional manifold generically breaks all conformal
supersymmetries. In the special case where the four-manifold is a
cone, we show that exactly two conformal supercharges remain
unbroken. We construct an off-shell formulation of the theory such
that the two unbroken conformal supercharges combine into a family
of topological charges parameterized by $ \mathbb{CP}^1$. The
resulting theory is topological in the sense that it is independent
of the metric on the three-dimensional base of the cone.
\end{abstract}
{\small
\begin{flushleft}
~~~~~~~~\textit{Keywords}: Topological field theory, super
Yang-Mills, superconformal symmetry.
\end{flushleft}}
\thispagestyle{empty}
\end{titlepage}
\newpage

\tableofcontents 
\section{Introduction}
Topological quantum field theory (TQFT) provides an intriguing link
between physics and mathematics. The study of TQFT was initiated
with \cite{Witten:1988ze}, wherein a quantum field-theoretical
representation of Donaldson invariants was given. A quantum field
theory is called topological if all vacuum expectation values (VEVs)
of a certain set of operators (`observables') are
metric-independent. In particular, { TQFTs of cohomological type}
are constructed as follows.  Let us assume there is a nilpotent
symmetry of the action $Q$, such that $Q^{2}=0$. It follows that, at
least formally, one can deform the Lagrangian by adding an arbitrary
$Q$-exact term without affecting the partition function or the VEVs
of observables (which are defined as elements in the cohomology of
$Q$). Since $Q$ is a symmetry of the action, the Lagrangian can be
expressed as a sum of a $Q$-exact and a $Q$-closed piece. The theory
is therefore independent of any coupling constant in the $Q$-exact
piece. Moreover, if the energy momentum tensor is $Q$-exact all VEVs
of observables are metric-independent and the theory is topological.

One way of constructing TQFTs is by `twisting' theories with
extended supersymmetry in Euclidean space. Roughly-speaking,
twisting can be thought of as an embedding of the rotation group in
the global $R$-symmetry group, thereby changing the spins of the
fields in the parent theory. For the resulting theory to be
topological on a general manifold, there should exist at least one
scalar among the twisted supersymmetry generators. The TQFT of
\cite{Witten:1988ze} was obtained, by twisting, from $\cN=2$
supersymmetric Yang-Mills theory in four dimensions. Similarly,
$\cN=4$  SYM in four dimensions can be twisted in three inequivalent
ways to obtain a TQFT \cite{ Vafa:1994tf,
Yamron:1988qc,Marcus:1995mq}. It has been subsequently  noted that
all three twisted theories can be thought of as world-volume
theories on D3-branes wrapping supersymmetric cycles in
string-theory compactifications \cite{bsv}. Two of the twists have
been extensively studied in \cite{Vafa:1994tf}, in connection to
instanton invariants. The third twist, which is the focus of the
present paper, has been studied in a series of papers
\cite{Marcus:1995mq,Blau:1996bx,
Labastida:1997vq,Labastida:1998sk,Lozano:1999ji,Unsal:2006qp}, and
has been recently found to be relevant to the geometric Langlands
program \cite{Kapustin:2006pk}.

Four-dimensional $\cN=4$ SYM in flat space is superconformal: in
addition to sixteen {ordinary} supercharges it possesses sixteen
{conformal}  supercharges. The aim  of the present paper is to
explore the effect of the twisting on the conformal supercharges. On
a general four-manifold all conformal supercharges will be  broken.
However, in the special case where the manifold is a cone over a
three-dimensional base, we find that the twisting leads to exactly
two unbroken conformal supercharges, $S_{0}$, $S_{5}$. We construct
an off-shell formulation of the theory such that the two twisted
conformal supercharges combine into a family of topological charges
$S_z:=z_{1}S_{0}+z_{2}S_{5}$, $z:=z_2/z_1$, parameterized by $
\mathbb{CP}^1$. In this formalism the components of the
energy-momentum tensor along the directions of the base of the cone
are manifestly $S_z$-exact -- since $S_z$ is nilpotent off-shell and
independent of the metric on the base of the cone. It follows that
(at least formally) the theory is topological in the sense that it
is independent of the metric on the base of the cone.

There are two important caveats in this statement. Firstly, as is
usually the case in the path-integral formulation of TQFTs, we shall
{assume} that the path-integral measure is well-defined and
$S_z$-invariant. Secondly, in order to define the functional
integration on a cone we have to impose boundary conditions on the
fields. As we explain in more detail in sections
\ref{subVanishingw}, \ref{subVanishings}, there is a certain freedom
in the choice of boundary conditions. Generic boundary conditions
will break superconformal invariance. Requiring that the scalar fields
vanish at infinity is  sufficient  for
the superconformal invariance to be preserved.

The aim of this paper is to furnish a new tool for the study of topological properties of
three-manifolds (viewed as bases of four-dimensional cones). We believe
that the topological off-shell theory we construct reveals a rich enough structure
to warrant further study. Of course the construction of the off-shell formalism should be
viewed as just
the first step; eventually we would like to identify the set of observables of the theory
 (for this, a more complete study of the boundary conditions will be necessary)
and their relevance to three-dimensional topological invariants. We will leave
this second nontrivial step for future investigation.

The organization of this paper is as follows. In section
\ref{SecPre} we  review some well-known facts about ten-dimensional
SYM and its reduction to $\cN=4$ SYM in four dimensions.
 In particular, we point out that in flat space the conformal
supersymmetries   can be viewed
as similarity transformations of the ordinary supersymmetries, with respect to
a certain inversion map.

Section \ref{SecTw} explains the  twisting of the $\cN=4$ super
Yang-Mills on which we focus  in this paper. We obtain the twisted
Lagrangian as well as the ordinary and conformal scalar
supersymmetry transformations. The superconformal algebra of the
twisted SYM in flat space contains two ordinary scalar supercharges
and  two conformal scalar supercharges. We show that the conformal
scalar supersymmetries survive on a curved four-manifold,  if and
only if the manifold is (locally) a {cone} over an arbitrary
three-dimensional base.
 We identify the most general  linear combination of   the four
scalar supercharges which is nilpotent,
and thus can be used to define a cohomological structure. Some important `vanishing
theorems' are presented in subsections \ref{subVanishingw} and
\ref{subVanishings}.

In section \ref{SecOff},
 we show that
there exists an off-shell formulation of the theory
such that the two twisted conformal supercharges combine into a
family of nilpotent operators
parameterized by $ \mathbb{CP}^1$. Moreover we show that (at least formally)
the theory is topological in the sense that it is independent of the metric
on the base of the cone.

 We conclude with a discussion of
open questions in section \ref{SecCoO}. The appendices contain some useful proofs and formul\ae{}.


\section{Review of super Yang-Mills\label{SecPre}}
In order to establish our conventions and notation, we now give a
brief review of super Yang-Mills in ten Euclidean dimensions and its
reduction to four-dimensional  $\cN=4$ super Yang-Mills.

By Wick-rotating the ordinary super Yang-Mills in ten dimensions, one obtains
a Euclidean  theory which is still formally supersymmetric in that it is invariant
under the same set of fermionic transformations as the original theory.
However, the minimal spinor representation in ten Euclidean dimensions
is no longer Majorana, resulting in a  complex action. Of course this phenomenon
is not new: it occurs routinely in string theory in a variety of different contexts.

There are essentially two ways to deal with this problem.
The point-of-view we take here is to simply ignore
the fact that the action is complex, as for example
in \cite{Kapustin:2006pk}.  This would naively seem to double
the fermionic degrees of freedom. However, the complex-conjugate
fermions do not appear in the Lagrangian. Therefore in the path-integral one
integrates over only half of the (complexified) fermionic degrees of freedom --
resulting in the same counting as for the Minkowskian theory.
Another way to state this is that the action is not required to be real --
it is only required to be holomorphic.
The other way to handle the Wick rotation is by imposing reality conditions
on the fields. These reality conditions are sometimes introduced in an ad hoc manner.
A systematic way to derive them is by time-like reduction of a higher-dimensional
theory, as demonstrated in \cite{blau}.

\subsection{Super Yang-Mills in ten dimensions}
Consider the $32\times 32$  gamma matrices in ten-dimensional Euclidean space
\be
\ba{ll}
\Gamma^{M}\Gamma^{N}+\Gamma^{N}\Gamma^{M}=2\delta^{MN}\,,
~~~~&~~~~1\leq M,N\leq 10\,.
\ea
\ee
These are Hermitian $\left(\Gamma^{M}\right)^{\dagger}=\Gamma^{M}$ and satisfy
\be
\ba{ll}
\left(\Gamma^{M}\right)^{T}=\left(\Gamma^{M}\right)^{\ast}
=\cC\Gamma^{M}\cC^{\dagger}\,,~~~~&~~~~
\cC=\cC^{\,T}=(\cC^{\dagger})^{-1}\,.
\ea
\label{gABC}
\ee
We define the ten-dimensional chirality operator by
\be
\Gamma^{(11)}:=-i\Gamma^{12\cdots 10}=(\Gamma^{(11)})^{\dagger}=(\Gamma^{(11)})^{-1}=
-\cC^{\dagger}(\Gamma^{(11)})^{T}\cC\,.
\label{g11}
\ee
The main Fierz  identity  is
\be
(\cC\Gamma^{M}P_{\pm})_{\halpha\hbeta}
(\cC\Gamma_{M}P_{\pm})_{\hgamma\hdelta}
+(\cC\Gamma^{M}P_{\pm})_{\hbeta\hgamma}
(\cC\Gamma_{M}P_{\pm})_{\halpha\hdelta}
+(\cC\Gamma^{M}P_{\pm})_{\hgamma\halpha}
(\cC\Gamma_{M}P_{\pm})_{\hbeta\hdelta}
 =0\,,
\label{Fierz10}
\ee
where $\halpha$, $\hbeta$, $\hgamma$, $\hdelta$ are the 32 component
spinorial indices, and $P_{\pm}$ are the
chiral/anti-chiral projection matrices
\be
\ba{ll}
P_{+}=\half(1+\Gamma^{(11)})\,,~~~~&~~~~P_{-}=\half(1-\Gamma^{(11)})\,.
\ea
\ee
Super Yang-Mills in ten Euclidean dimensions is given by:
\be
\cL_{{10D}}
=\tr\Big[\textstyle{\frac{1}{4}}F_{MN}F^{MN}+\half\bar{\Psi}\Gamma^{M}D_{M}\Psi
\Big]\,,
\ee
where the gaugino  $\Psi$ is a chiral spinor
\be
\Gamma^{(11)}\Psi=+\Psi\,,
\label{chiralPsi}
\ee
and $\bar{\Psi}$ is defined by\footnote{Note that although the gaugino is complex in ten Euclidean dimensions,
its complex conjugate does not appear in the Lagrangian.}
\be
\ba{ll}
\bar{\Psi}:=\Psi^{T}\cC\,,~~~~&~~~~
\bar{\Psi}\Gamma^{(11)}=-\bar{\Psi}\,.
\ea
\ee
In our conventions the gauge field
strength and the gauge-covariant derivative read
\be
\ba{ll}
F_{MN}=\partial_{M}A_{N}-\partial_{N}A_{M}-i[A_{M},A_{N}]\,,~~&~~D_{M}\Psi=
\nabla_{M}\Psi-i[A_{M},\Psi]\,,
\ea
\ee
so that  gauge transformations are
\be
\ba{lll}
A_{M}\rightarrow gA_{M}g^{-1}+ig\partial_{M}g^{-1}\,,~~&~~
F_{MN}\rightarrow gF_{MN}g^{-1}\,,~~&~~\Psi\rightarrow g\Psi g^{-1}\,.
\ea
\ee
The sixteen {ordinary supersymmetries} are given by
\be
\ba{ll}
\delta A_{M}=\bar{\Psi}\Gamma_{M}\varepsilon_{+}=-\bar{\varepsilon}_{+}
\Gamma_{M}\Psi\,,~~~~&~~~~
\delta\Psi=\half F_{MN}\Gamma^{MN}\varepsilon_{+}\,,
\ea
\label{ordinarySUSY}
\ee
where the supersymmetry parameter $\varepsilon_{+}$ is a constant Weyl spinor, and
the subscript denotes positive chirality in ten dimensions.

\subsection{Dimensional reduction and superconformal symmetry}
We now consider the dimensional reduction to
 four-dimensional  $\cN=4$ super Yang-Mills. Let us denote  the
four-dimensional  coordinates by $x^{\mu}$, $\,\mu=1,2,3,4$. We also
set $\Phi_{I}:= A_{I}$, $\,5\leq I\leq 10$. Four-dimensional
$\cN=4$ super Yang-Mills is then given by \be \cL_{{4D}}
=\tr\Big[\textstyle{\frac{1}{4}}F_{\mu\nu}F^{\mu\nu}+\half
D_{\mu}\Phi_{I}D^{\mu}\Phi^{I}
-\textstyle{\frac{1}{4}}\left[\Phi_{I},\Phi_{J}\right]\left[\Phi^{I},\Phi^{J}\right]
+\half\bar{\Psi}\Gamma^{\mu}D_{\mu}\Psi-i\half\bar{\Psi}\Gamma^{I}\left[\Phi_{I},
\Psi\right] \Big]\,. \label{N=4SYM} \ee In addition to the sixteen
ordinary supersymmetries (\ref{ordinarySUSY}), there are sixteen
{conformal supersymmetries} \be \ba{ll} \delta
A_{M}=\bar{\Psi}\Gamma_{M}X\varepsilon_{-}=-\bar{\varepsilon}_{-}X\Gamma_{M}\Psi\,,
~~~~&~~~~ \delta\Psi=\left(\half
F_{MN}\Gamma^{MN}X-2\Phi_{I}\Gamma^{I}\right)\varepsilon_{-}\,, \ea
\label{conformalSUSY} \ee where $X:=x^{\mu}\Gamma_{\mu}$ and the
conformal supersymmetry parameter  $\varepsilon_{-}$ is a Weyl
spinor of negative chirality,
$\Gamma^{(11)}\varepsilon_{-}=-\varepsilon_{-}$.\footnote{For
simplicity the spinors are given in ten-dimensional notation.}

Both ordinary and conformal supersymmetries  can be collected
in terms of a single 32-component spinor
\be
\ba{ll}
\varepsilon=\varepsilon_{+}+\varepsilon_{-}\,,~~~~&~~~~   \varepsilon_{\pm}
=P_{\pm}\varepsilon\,.
\ea
\label{32xi}
\ee
All 32 supersymmetries are then given by
\be
\ba{l}
\delta A_{M}=\bar{\Psi}\Gamma_{M}(1+x^{\mu}\Gamma_{\mu})\varepsilon
=-\bar{\varepsilon}(1+x^{\mu}\Gamma_{\mu})\Gamma_{M}\Psi\,,\\
{}\\
\delta\Psi=P_{+}
\left[\half F_{MN}\Gamma^{MN}(1+x^{\mu}\Gamma_{\mu})-2\Phi_{I}\Gamma^{I}\right]
\varepsilon\,,
\ea
\ee
which implies
\be
\delta\bar{\Psi}=\bar{\varepsilon}
\left[-\half (1+x^{\mu}\Gamma_{\mu}) F_{MN}\Gamma^{MN}-2\Phi_{I}\Gamma^{I}\right]P_{-}\,.
\ee
Accordingly, the Lagrangian (\ref{N=4SYM}) transforms
into  a total derivative
\be
\delta\cL_{{4D}}=
\partial_{\mu}\tr\Big[F^{\mu N}\delta
A_{N}-\half\bar{\Psi}\Gamma^{\mu}\delta\Psi\Big]\,,
\ee
leaving the action invariant.

{}From the corresponding Noether current
\be
J^{\mu}=\tr\!\left(\bar{\Psi}\Gamma^{\mu}\delta\Psi\right)=
-\tr\!\left(\delta\bar{\Psi}\Gamma^{\mu}\Psi\right)\,,
\ee
we obtain a 32-component supercurrent
\be
\ba{l}
\cQ^{\mu}=\tr\!\left[\left(\half (1+x^{\nu}\Gamma_{\nu}) F_{KL}\Gamma^{KL}
+2\Phi_{I}\Gamma^{I}\right)\Gamma^{\mu}\Psi\right]\,,\\
{}\\
\bar{\cQ}^{\mu}= \tr\!\left[\bar{\Psi}\Gamma^{\mu}\left(-\half
F_{KL}\Gamma^{KL}(1+x^{\nu}\Gamma_{\nu})+2\Phi_{I}\Gamma^{I}\right)\right]
=(\cQ^{\mu})^{T}\cC\,,
\ea
\ee
where we have set
\be
J^{\mu}:=-\bar{\cQ}^{\mu}\varepsilon=+\bar{\varepsilon}\cQ^{\mu}\,.
\label{JQx}
\ee
The chiral  and anti-chiral parts  $P_{+}\cQ^{\mu}$, $P_{-}\cQ^{\mu}$,
correspond to the
conformal and ordinary supersymmetries  respectively.

It is worthwhile  to note the existence of an inversion map $\cI$
defined in flat background~\cite{Osborn:1993cr,JHP4D,PhDPark}
\be
\ba{cll}
x^{\mu}~&~\stackrel{\cI}{-\!\!\!-\!\!\!-\!\!\!\longrightarrow}~&~x^{\prime}{}^{\mu}=
\displaystyle{\frac{x^{\mu}}{x^{2}}\,,}\\
{}&{}&{}\\
A_{\mu}(x)~&~\stackrel{\cI}{-\!\!\!-\!\!\!-\!\!\!\longrightarrow}
~&~A^{\prime}_{\mu}(x)=
\displaystyle{\frac{1}{x^{2}}
\left(\delta_{\mu}{}^{\nu}-2\frac{x_{\mu}x^{\nu}}{x^{2}}\right)A_{\nu}(x^{\prime})\,,}\\
{}&{}&{}\\
\Phi_{J}(x)~&~\stackrel{\cI}{-\!\!\!-\!\!\!-\!\!\!\longrightarrow}
~&~\Phi^{\prime}_{J}(x)=
\displaystyle{\frac{1}{x^{2}}\Phi_{J}(x^{\prime})\,,}\\
{}&{}&{}\\
\Psi(x)~~&~\stackrel{\cI}{-\!\!\!-\!\!\!-\!\!\!\longrightarrow}~&~\Psi^{\prime}(x)=
\displaystyle{i\frac{x^{\mu}\Gamma_{\mu}}{\left(x^{2}\right)^{2}}\Psi(x^{\prime})\,.}
\ea \label{inversion}
\ee
The inversion map   is an involution:
$\cI=\cI^{-1}$. The four-dimensional $\cN=4$ super Yang-Mills
action~(\ref{N=4SYM}) is invariant under the action of  $\cI$. The
ordinary and conformal supersymmetry transformations,
(\ref{ordinarySUSY}) and  (\ref{conformalSUSY}), are  related to
each other by the similarity transformation~\cite{JHP4D}:
\be
\mathrm{conformal~supersymmetry}=\cI\circ\left(\,
\mathrm{ordinary~supersymmetry}\,\right)\circ\cI\,.
\label{inversionSQ}
\ee
The inversion map flips the chirality of the
fermion, which is consistent with the fact that the two
supersymmetry parameters $\varepsilon_{+}$ and $\varepsilon_{-}$
have opposite chiralities. Note, however,
that this symmetry  is broken in a generic curved background.

\section{Twisted    $\cN=4$ super Yang-Mills in four dimensions\label{SecTw}}
We now come to the description of the
  twist. In the following subsections, we also
discuss the unbroken conformal supercharges and derive certain
important `vanishing theorems'.
\subsection{Description of the twist}
Under ${Spin}(10)\rightarrow Spin(2)\times Spin(4)\times{}Spin(4)$,
the ten-dimensional gamma matrices can be decomposed as
\be
\ba{ll}
\Gamma^{\mu}=\tau^{1}\otimes\gamma^{\mu}\otimes 1\,,~~~~&~~~~~
\Gamma^{\mu+4}=\tau^{2}\otimes 1\otimes\gamma^{\mu}\,,\\
{}&{}\\
\Gamma^{9}=\tau^{1}\otimes\gamma^{(5)}\otimes 1\,,~~~~&~~~~~
\Gamma^{10}=\tau^{2}\otimes 1\otimes\gamma^{(5)}\,,
\ea
\label{forG}
\ee
where $\tau^{i}$, $i=1,2,3$,
are $2\times 2$ Pauli matrices
\be
\ba{lll}
\tau^{1}=\left(\ba{cc}0&~1\\1&~0\ea\right)\,,~~~~&~~~~
\tau^{2}=\left(\ba{cc}0&-i\\+i&0\ea\right)\,,~~~~&~~~~
\tau^{3}=\left(\ba{cc}+1&\,0\\0&-1\ea\right)\,,
\ea
\ee
and $\gamma^{\mu}$, $\mu=1,2,3,4$ are four-dimensional gamma-matrices
\be
\ba{ll}
\gamma^{\mu}\gamma^{\nu}+\gamma^{\nu}\gamma^{\mu}=2\delta^{\mu\nu}\,,~~~~&~~~~
\left(\gamma^{\mu}\right)^{\dagger}=\gamma^{\mu}\,,\\
{}&{}\\
\left(\gamma^{\mu}\right)^{T}=C\gamma^{\mu}C^{-1}\,,~~~~&~~~~
C=-C^{\,T}=\left(C^{\dagger}\right)^{-1}\,,\\
{}&{}\\
\gamma^{(5)}:=\gamma^{1}\gamma^{2}\gamma^{3}\gamma^{4}\,,~~~~&~~~~
\left(\gamma^{(5)}\right)^{T}=C\gamma^{(5)}C^{-1}\,.
\ea
\ee
Taking (\ref{gABC}) and (\ref{g11}) into account, it follows that
\be
\ba{ll}
\cC=\tau^{1}\otimes C\otimes C \,,~~~~~&~~~~~
\Gamma^{(11)}=\tau^{3}\otimes 1\otimes 1\,,
\ea
\label{CtCC}
\ee
so that the ten-dimensional chirality
coincides with the chirality in the $Spin(2)$ part.

The fermions $\Psi_{\pm\alpha\beta}$  carry three indices for
${}Spin(2)\times{}Spin(4)\times{}Spin(4)$. The focus of the present paper is
the twist considered in
\cite{Yamron:1988qc,Marcus:1995mq,Blau:1996bx,
Labastida:1997vq,Labastida:1998sk,Lozano:1999ji,Kapustin:2006pk}: it amounts to
replacing the four-dimensional rotation group by the diagonal subgroup of
$Spin(4)\times{}Spin(4)$.
Accordingly, the twisted ten-dimensional chiral fermion~(\ref{chiralPsi})
admits the following expansion
\be
\ba{ll}
\Psi_{+}=\half\Big(\eta +\psi_{\mu}\gamma^{\mu}
+\half\chi_{\mu\nu}\gamma^{\mu\nu}+\omega_{\mu}\gamma^{\mu}\gamma^{(5)}
+\zeta\gamma^{(5)}\Big)C^{-1}\,,~~~~&~~~~\Psi_{-}=0\,.
\ea
\label{forF}
\ee
It follows that the fermions
decompose into a pair of anticommuting scalars  $\eta$, $\zeta$,
a pair of vectors $\psi_{\mu}$, $\omega_{\mu}$ and a two-form
$\chi_{\mu\nu}=-\chi_{\nu\mu}$. The chiral  and  anti-chiral
supersymmetry parameters $\varepsilon_{+}$, $\varepsilon_{-}$ in (\ref{32xi}) admit
similar expansions,
as we shall see later  in (\ref{forxi}).
In particular, there will be two ordinary scalar  and two conformal scalar
supercharges.

For the bosons we set
\be
\ba{ll}
A^{\pm}_{\mu}:=A_{\mu}\pm i\phi_{\mu}\,,~~~~&~~~~\phi_{\mu}:=\Phi_{4+\mu}\,,\\
{}&{}\\
\varphi:=\Phi_{9}+i\Phi_{10}\,,~~~~&~~~~
\bvarphi:=\Phi_{9}-i\Phi_{10}\,.
\ea
\label{forB}
\ee
These obey the following  reality conditions:
\be
\ba{lll}
\left(A^{+}_{\mu}\right)^{\dagger}=A^{-}_{\mu}\,,~~~~&~~~~
\left(\phi_{\mu}\right)^{\dagger}=\phi_{\mu}\,,~~~~&~~~~
\left(\varphi\right)^{\dagger}=\bvarphi\,,
\ea
\label{reality}
\ee
while there is  no analogous  constraint  for fermions\footnote{
If we started with super Yang-Mills in ten-dimensional Minkowski
instead of Euclidean spacetime, the fermions
would have to satisfy the  Majorana condition.
This  would result in
reality conditions for $\eta,\zeta,\psi_{\mu},\omega_{\mu},\chi_{\mu\nu}$.}.

\subsection{Twisted Lagrangian}
Taking  (\ref{forG}), (\ref{forF}), (\ref{forB}) into account, it is
straightforward to rewrite the $\cN=4$ super Yang-Mills (\ref{N=4SYM})
in terms of the anticommuting fields
$\eta,\zeta,\psi_{\mu},\omega_{\mu},\chi_{\mu\nu}$ and the bosons
$A^{\pm}_{\mu}$, $\phi_{\mu}$, $\varphi$, $\bvarphi$.
The resulting action defines our  twisted $\cN=4$
super Yang-Mills in four-dimensions:\footnote{
N. Nekrasov has pointed out that
this action can be obtained by dimensional reduction of five-dimensional twisted
super Yang-Mills~\cite{Nikita-Park}.}
\be
\ba{ll}
\cS_{\mathrm{twisted}}=\displaystyle{\int{\rm d^{4}}x\,
\,\cL_{\mathrm{twisted}}\,,}~~~~&~~~~\displaystyle{
\cL_{\mathrm{twisted}}=\cL_{\mathrm{top}}+\sqrt{g}\,L_{g}\,,}
\ea
\label{Stwisted}
\ee
where
\be
\ba{l}
\cL_{\mathrm{top}}=\epsilon^{\kappa\lambda\mu\nu}\,
\tr\Big[\,\half\omega_{\kappa}\cD^{+}_{\lambda}\chi_{\mu\nu}
-i\textstyle{\frac{1}{8}}\varphi\left\{\chi_{\kappa\lambda},\chi_{\mu\nu}\right\}
\Big]\,,\\
{}\\
\ba{ll}
L_{g}=\tr\Big[&
\textstyle{\frac{1}{4}}F^{+}_{\mu\nu}F^{-\mu\nu}
-\half h^{2}+h\cD_{\mu}\phi^{\mu}+\textstyle{\frac{1}{4}}\cD^{+}_{\mu}\varphi
\cD^{-\mu}\bvarphi+\textstyle{\frac{1}{4}}\cD^{-}_{\mu}\varphi\cD^{+\mu}\bvarphi
+\textstyle{\frac{1}{8}}\left[\varphi,\bvarphi\right]^{2}\\
{}&{}\\
{}&\,+\,{\eta}\cD^{+}_{\mu}\psi^{\mu}+{\zeta}\cD^{-}_{\mu}\omega^{\mu}
+\chi^{\mu\nu}\cD^{-}_{\mu}\psi_{\nu}+i\varphi\left\{\eta,\zeta\right\}
-i\bvarphi\left\{\psi_{\mu},\omega^{\mu}\right\}\Big]\,. \ea \ea
\label{topaction} \ee In the above, we have introduced  an auxiliary
bosonic scalar $h$. We have coupled the action to a generic curved
background with metric $g_{\mu\nu}$.\footnote{
For the coupling of the topological field theory to gravity see \textit{e.g.}
\cite{Karlhede:1988ax}.}  Moreover,
$\epsilon^{\kappa\lambda\mu\nu}$ is the totally antisymmetric tensor
density, such that  $\epsilon^{1234}=1$ and $\epsilon_{1234}=g$. All
derivatives are covariant with respect to both  diffeomorphisms and
gauge transformations: \be \ba{ll} \cD_{\mu}\phi^{\nu}=
\nabla_{\mu}\phi^{\nu}-i[A_{\mu},\phi^{\nu}]\,,~~~~&~~~~
\cD_{\mu}^{\pm}{\eta}=\nabla_{\mu}{\eta}-i[A_{\mu}^{\pm},{\eta}]\,,\\
{}&{}\\
\cD_{\mu}^{\pm}\psi_{\nu}=\nabla_{\mu}\psi_{\nu}
-i[A^{\pm}_{\mu},\psi_{\nu}]\,,~~~~&~~~~
\cD_{\mu}^{\pm}\chi_{\nu\lambda}=\nabla_{\mu}\chi_{\nu\lambda}
-i[A^{\pm}_{\mu},\chi_{\nu\lambda}]\,.
\ea
\ee
Note that $\cL_{\mathrm{top}}$ is manifestly metric-independent, as follows from the
fact that the Christoffel connection is torsion-free.

Moreover, we have defined the  field strengths
\be
\ba{l}
F^{\pm}_{\mu\nu}:=\partial_{\mu}A^{\pm}_{\nu}-\partial_{\nu}A^{\pm}_{\mu}
-i[A^{\pm}_{\mu},A^{\pm}_{\nu}]=F_{\mu\nu}+ i[\phi_{\mu},\phi_{\nu}]
\pm i\left(\cD_{\mu}\phi_{\nu}-\cD_{\nu}\phi_{\mu}\right)\,.
\ea
\label{FFF}
\ee
Note that the superscript $\pm$ above does \textit{not} denote (anti) self-duality.
It is also useful to define a  covariant tensor
$\cH_{\mu\nu}$ by
\be
\cH_{\mu\nu}:
=\nabla_{\mu}A^{-}_{\nu}-\nabla_{\nu}A^{+}_{\mu}-i[A^{+}_{\mu},A^{-}_{\nu}]\,,
\ee
such that
\be
\cH_{\mu\nu}=
F_{\mu\nu}-i[\phi_{\mu},\phi_{\nu}]-i\cD_{\mu}\phi_{\nu}-i\cD_{\nu}\phi_{\mu}
=F^{+}_{\mu\nu}-2i\cD^{+}_{\mu}\phi_{\nu}
=F^{-}_{\mu\nu}-2i\cD^{-}_{\nu}\phi_{\mu}\,,
\ee
and
\be
\cD_{\mu}\phi^{\mu}=\half i\,\cH_{\mu}{}^{\mu}\,.
\ee
The field strengths can be seen to obey the following generalized Bianchi identities
\be
\ba{ll}
\cD^{+}_{\lambda}F^{-}_{\mu\nu}+\cD^{-}_{\nu}\cH_{\lambda\mu}
-\cD^{-}_{\mu}\cH_{\lambda\nu}=0\,,~~~~&~~~~
{}\cD^{-}_{\lambda}F^{+}_{\mu\nu}+\cD^{+}_{\mu}\cH_{\nu\lambda}
-\cD^{+}_{\nu}\cH_{\mu\lambda}=0\,,\\
{}&{}\\
{}\cD^{+}_{\lambda}F^{+}_{\mu\nu}
+\cD^{+}_{\mu}F^{+}_{\nu\lambda}+\cD^{+}_{\nu}F^{+}_{\lambda\mu}=0\,,~~~~&~~~~
{}\cD^{-}_{\lambda}F^{-}_{\mu\nu}
+\cD^{-}_{\mu}F^{-}_{\nu\lambda}+\cD^{-}_{\nu}F^{-}_{\lambda\mu}=0\,.
\ea
\label{Bianchi}
\ee
It is important to note that the identities above are valid
on any curved manifold.
Moreover, taking the following identities  into account\footnote{
In our conventions the Riemann tensor is given by
 $R^{\lambda}{}_{\kappa\mu\nu}:=\partial_{\mu}\Gamma^{\lambda}_{\nu\kappa}-
\partial_{\nu}\Gamma^{\lambda}_{\mu\kappa}
+\Gamma^{\lambda}_{\mu\sigma}\Gamma^{\sigma}_{\nu\kappa}-
\Gamma^{\lambda}_{\nu\sigma}\Gamma^{\sigma}_{\mu\kappa}$.}
\be
\ba{ll}
{}[\cD_{\mu},\cD_{\nu}]\phi_{\lambda}
+R^{\rho}{}_{\lambda\mu\nu}\phi_{\rho}+i[F_{\mu\nu},\phi_{\lambda}]=0\,,~~~~&~~~~
{}[\cD_{\mu},\cD_{\nu}]\phi^{\nu}+R_{\mu\nu}\phi^{\nu}+i[F_{\mu\nu},\phi^{\nu}]=0\,,
\ea
\label{DDRF}
\ee
the bosonic part of the Lagrangian  containing
$F_{\mu\nu}$ and $\phi_{\mu}$
can be rewritten as
\be
\ba{ll}
\tr\Big(
\textstyle{\frac{1}{4}}F^{+}_{\mu\nu}F^{-\mu\nu}
+&\half \left(\cD_{\mu}\phi^{\mu}\right)^{2}\Big)=
\nabla_{\mu}
\tr\Big(\half\phi^{\mu}\cD^{\nu}\phi_{\nu}-\half
\phi^{\nu}\cD_{\nu}\phi^{\mu}\Big)\\
{}&{}\\
&+\tr\Big(\textstyle{\frac{1}{4}}F_{\mu\nu}F^{\mu\nu}
+\half\cD_{\mu}\phi_{\nu}\cD^{\mu}\phi^{\nu}-\textstyle{\frac{1}{4}}
[\phi_{\mu},\phi_{\nu}][\phi^{\mu},\phi^{\nu}]+
\half R_{\mu\nu}\phi^{\mu}\phi^{\nu}\Big)\,.
\ea
\label{RicciA}
\ee
The Ricci term on the right-hand side is necessary to ensure the invariance of the action under the
unbroken scalar supersymmetries.

The equations of motion are
\be
\ba{l}
\cD^{+}_{\nu}F^{-\mu\nu}+i\cD^{-\mu}h-i\half\left[\varphi,\cD^{-\mu}\bvarphi\right]
-i\half\left[\bvarphi,\cD^{-\mu}\varphi\right]
+2i\left\{{\eta},\psi^{\mu}\right\}-i\displaystyle{\frac{1}{\sqrt{g}}}\,
\epsilon^{\mu\nu\kappa\lambda}\left\{\omega_{\nu},\chi_{\kappa\lambda}\right\}=0\,,\\
{}\\
\cD^{-}_{\nu}F^{+\mu\nu}-i\cD^{+\mu}h-i\half\left[\varphi,\cD^{+\mu}\bvarphi\right]
-i\half\left[\bvarphi,\cD^{+\mu}\varphi\right]
+2i\left\{\zeta,\omega^{\mu}\right\}
+2i\left\{\psi_{\nu},\chi^{\mu\nu}\right\}=0\,,\\
{}\\
\cD^{+}_{\mu}\cD^{-\mu}\varphi+\cD^{-}_{\mu}\cD^{+\mu}\varphi+\left[\varphi,\left[
\varphi,\bvarphi\right]\right]+4i\left\{\psi_{\mu},\omega^{\mu}\right\}=0\,,\\
{}\\
\cD^{+}_{\mu}\cD^{-\mu}\bvarphi+\cD^{-}_{\mu}\cD^{+\mu}\bvarphi-\left[\bvarphi,\left[
\varphi,\bvarphi\right]\right]-4i\left\{\eta,\zeta\right\}+i
\displaystyle{\frac{1}{\,2\sqrt{g}}}\,
\epsilon^{\kappa\lambda\mu\nu}\left\{\chi_{\kappa\lambda},\chi_{\mu\nu}\right\}=0\,,\\
{}\\
h-\cD_{\mu}\phi^{\mu}=0\,,\\
{}\\
\cD^{-\mu}\psi^{\nu}-\cD^{-\nu}\psi^{\mu}-\displaystyle{\frac{1}{\sqrt{g}}}\,
\epsilon^{\mu\nu\kappa\lambda}\left(
\cD^{+}_{\kappa}\omega_{\lambda}+i\half\left[\chi_{\kappa\lambda},\varphi\right]\right)
=0\,,\\
{}\\
\cD^{+}_{\mu}{\eta}-\cD^{-\nu}\chi_{\mu\nu}-i\left[\omega_{\mu},\bvarphi\right]=0\,,\\
{}\\
\cD^{-}_{\mu}\zeta-i\left[\psi_{\mu},\bvarphi\right]+
\displaystyle{\frac{1}{\,2\sqrt{g}}}\,
\epsilon_{\mu}{}^{\nu\kappa\lambda}\cD^{+}_{\nu}\chi_{\kappa\lambda}=0\,,\\
{}\\
\cD^{+}_{\mu}\psi^{\mu}+i\left[\zeta,\varphi\right]=0\,,\\
{}\\
\cD^{-}_{\mu}\omega^{\mu}+i\left[\eta,\varphi\right]=0\,.
\ea
\label{EOM}
\ee
%
%
Finally, the energy-momentum tensor is given by\footnote{In deriving (\ref{EMT}),
one has to take into account the variation of the Christoffel connection
in the covariant derivative $\nabla_{\mu}$.}
\be
\ba{ll}
T^{\mu\nu}&=\displaystyle{\frac{2}{\sqrt{g}}\,
\frac{\,\delta\cS_{\mathrm{twisted}}\,}{\delta g_{\mu\nu}}\,,}\\
{}&{}\\
{}&\ba{ll}
=g^{\mu\nu}
\tr\!\Big[&
\textstyle{\frac{1}{4}}F^{+}_{\kappa\lambda}F^{-\kappa\lambda}
-\half h^{2}-\phi^{\kappa}\cD_{\kappa}h
+\textstyle{\frac{1}{4}}\cD^{+}_{\kappa}\varphi
\cD^{-\kappa}\bvarphi+\textstyle{\frac{1}{4}}\cD^{-}_{\kappa}\varphi
\cD^{+\kappa}\bvarphi+\textstyle{\frac{1}{8}}\left[\varphi,\bvarphi\right]^{2}\\
{}&{}\\
{}&\,+\psi^{\kappa}\cD^{+}_{\kappa}{\eta}+\omega^{\kappa}\cD^{-}_{\kappa}{\zeta}
+\chi^{\kappa\lambda}\cD^{-}_{\kappa}\psi_{\lambda}+i\varphi\left\{\eta,\zeta\right\}
-i\bvarphi\left\{\psi_{\mu},\omega^{\mu}\right\}\Big]\ea\\
{}&{}\\
{}&\,\ba{ll}
+\tr\!\Big[&-F^{+(\mu}{}_{\kappa}F^{-\nu)\kappa}+2\phi^{(\mu}\cD^{\nu)}h
-\half\cD^{+(\mu}\varphi\cD^{-\nu)}\bvarphi
-\half\cD^{-(\mu}\varphi\cD^{+\nu)}\bvarphi\\
{}&{}\\
{}&\,
-2\psi^{(\mu}\cD^{+\nu)}{\eta}-2\omega^{(\mu}\cD^{-\nu)}{\zeta}
+2\chi^{\kappa(\mu}\cD^{-\nu)}\psi_{\kappa}
-2\chi^{\kappa(\mu}\cD^{-}_{\kappa}\psi^{\nu)}
+2i\bvarphi\left\{\psi^{(\mu},\omega^{\nu)}\right\}\Big]\,,\ea
\ea
\label{EMT}
\ee
where the brackets denote symmetrization with weight one.
%
\subsection{Superconformal symmetry}
Upon twisting, as in the case of the Yang-Mills gaugino~(\ref{forF}),
the chiral ordinary  supersymmetry and the anti-chiral conformal supersymmetry parameters,
$\varepsilon_{+}$ and  $\varepsilon_{-}$ respectively,
decompose as
\be
\ba{l}
\varepsilon_{+}=\half\Big(\varepsilon_{0} +\varepsilon_{\mu}\gamma^{\mu}
+\half \varepsilon_{\mu\nu}\gamma^{\mu\nu}+{\varepsilon}^{\prime}_{\mu}
\gamma^{\mu}\gamma^{(5)}
+\varepsilon_{5}\gamma^{(5)}\Big)C^{-1}\,,\\
{}\\
\varepsilon_{-}=\half\Big(\tilde{\varepsilon}_{0} +
\tilde{\varepsilon}_{\mu}\gamma^{\mu}
+\half \tilde{\varepsilon}_{\mu\nu}\gamma^{\mu\nu}+
\tilde{\varepsilon}^{\prime}_{\mu}\gamma^{\mu}\gamma^{(5)}
+\tilde{\varepsilon}_{5}\gamma^{(5)}\Big)C^{-1}\,.
\ea
\label{forxi}
\ee
The corresponding supercharges $Q_{-}$,  $S_{+}$
have the opposite chiralities from   (\ref{JQx}), (\ref{CtCC}),  and decompose similarly
\be
\ba{l}
Q_{-}=\half\Big(Q_{0} +Q_{\mu}\gamma^{\mu}
+\half Q_{\mu\nu}\gamma^{\mu\nu}+\tilde{Q}_{\mu}\gamma^{\mu}\gamma^{(5)}
+Q_{5}\gamma^{(5)}\Big)C^{-1}\,,\\
{}\\
S_{+}=\half\Big(S_{0} +S_{\mu}\gamma^{\mu}
+\half S_{\mu\nu}\gamma^{\mu\nu}+\tilde{S}_{\mu}\gamma^{\mu}\gamma^{(5)}
+S_{5}\gamma^{(5)}\Big)C^{-1}\,.
\ea
\label{forQS}
\ee
Altogether, there are $\left(1+4+6+4+1\right)\times 2=32$ components, all of which
are unbroken in a flat background.
In  curved backgrounds, however, in order to have unbroken supersymmetries
it is necessary that the corresponding supersymmetry parameters should be
covariantly constant.
Generically, this requirement can only be met for  the scalar parameters
$\varepsilon_{0},\,\varepsilon_{5},\,\tilde{\varepsilon}_{0},\,\tilde{\varepsilon}_{5}$.
Hence, in a generic curved background
all the non-scalar supersymmetries $Q_{\mu},\tilde{Q}_{\mu},Q_{\mu\nu},
S_{\mu},\tilde{S}_{\mu},S_{\mu\nu}$ are   broken\footnote{Manifolds
which admit  covariantly-constant tensors are restricted to
be special-holonomy.}.

In addition, one has to check that in putting the
theory on a curved background the action remains supersymmetric. This is not obvious: in
passing from flat to curved background
there may be curvature terms arising from the commutator of two covariant
derivatives $[\nabla_\mu,\nabla_\nu]$, spoiling the invariance.
In the case at hand, one can  verify  that
the two ordinary scalar supercharges  $Q_{0}$ and $Q_{5}$ of
(\ref{Qtransf}) indeed give rise to
the two unbroken topological symmetries of the twisted super
Yang-Mills~(\ref{topaction}), as shown  in \cite{Yamron:1988qc,Marcus:1995mq,
Blau:1996bx,Labastida:1997vq,Labastida:1998sk,Lozano:1999ji,Kapustin:2006pk}.
%
%
Explicitly the action of $Q_{0}$, $Q_{5}$ is given by:
\be
\ba{ll}
{}\left[Q_{0},A^{+}_{\mu}\right]=0\,,~~~~&~~~~
\left[Q_{5},A^{+}_{\mu}\right]=-2\omega_{\mu}\,,\\
{}&{}\\
{}\left[Q_{0},A^{-}_{\mu}\right]=-2\psi_{\mu}\,,~~~~&~~~~
\left[Q_{5},A^{-}_{\mu}\right]=0\,,\\
{}&{}\\
{}\left[Q_{0},\varphi\right]=0\,,~~~~&~~~~
\left[Q_{5},\varphi\right]=0\,,\\
{}&{}\\
{}\left[Q_{0},\bvarphi\right]=-2\zeta\,,~~~~&~~~~
\left[Q_{5},\bvarphi\right]=-2\eta\,,\\
{}&{}\\
{}\left[Q_{0},h\right]=\left[\varphi,\zeta\right]\,,~~~~&~~~~
\left[Q_{5},h\right]=-\left[\varphi,\eta\right]\,,\\
{}&{}\\
{}\left\{Q_{0},\chi_{\mu\nu}\right\}=F^{+}_{\mu\nu}\,,~~~~&~~~~
\left\{Q_{5},\chi_{\mu\nu}\right\}
=\displaystyle{-\frac{1}{\,2\sqrt{g}\,}
\epsilon_{\mu\nu}{}^{\,\kappa\lambda}F^{-}_{\kappa\lambda}
=-\left(\star\,F^{-}\right)_{\mu\nu}\,,}\\
{}&{}\\
{}\left\{Q_{0},\psi_{\mu}\right\}=0\,,~~~~&~~~~
\left\{Q_{5},\psi_{\mu}\right\}=\cD^{-}_{\mu}\varphi\,,\\
{}&{}\\
{}\left\{Q_{0},\omega_{\mu}\right\}=\cD^{+}_{\mu}\varphi\,,~~~~&~~~~
\left\{Q_{5},\omega_{\mu}\right\}=0\,,\\
{}&{}\\
{}\left\{Q_{0},\eta\right\}=ih+i\half\left[\varphi,\bvarphi\right]\,,~~~~&~~~~
\left\{Q_{5},\eta\right\}=0\,,\\
{}&{}\\
{}\left\{Q_{0},\zeta\right\}=0\,,~~~~&~~~~
\left\{Q_{5},\zeta\right\}=-ih+i\half\left[\varphi,\bvarphi\right]\,,\\
{}&{}\\
{}\left[Q_{0},g_{\mu\nu}\right]=0\,,~~~~&~~~~
\left[Q_{5},g_{\mu\nu}\right]=0\,.
\ea
\label{Qtransf}
\ee
In contrast to the above ordinary scalar supercharges,
the conformal scalar supercharges $S_{0}$ and $S_{5}$
are not straightforwardly
realized in the twisted super Yang-Mills as Noether symmetries,
because the conformal supersymmetry transformations in flat space
involve the space coordinate $x^{\mu}$ explicitly,  see (\ref{conformalSUSY}).
In a curved background, $x^{\mu}$ should be replaced by a vector field $v^{\mu}$
satisfying the following relation:\footnote{
See \cite{deMedeiros:2001kx} for a different treatment.}
\be
\displaystyle{\nabla_{\mu}v^{\nu}=\delta_{\mu}{}^{\nu}\,.}
\label{conforv}
\ee
Indeed, the condition  $\partial_{\mu}x^{\nu}=\delta_{\mu}{}^{\nu}$ is the only
property of $x^{\mu}$ which is required for the invariance of the action
in a flat background. Moreover, it can be checked explicitly that the condition
(\ref{conforv}) suffices for the conformal scalar supercharges $S_{0}$ and $S_{5}$
to survive as symmetries of the action in a curved background,
provided the  vector field $v^{\mu}$ exists.
It is straightforward to see that any cone
\be
\ba{lll}
{\rm{d}}s^{2}={\rm{d}}r^{2}+r^{2}{\rm{d}}\hat{s}^{2}\,,~~~~&~~~~
{\rm{d}}\hat{s}^{2}=\hat{g}_{ij}(y){\rm{d}}y^{i}{\rm{d}}y^{j}\,,~~~~&~~~~
i,j=1,2,3\,,
\ea
\label{cone}
\ee
admits such a vector field:
\be
\displaystyle{v=r\frac{\,\partial~}{\,\partial r}\,.}
\ee
Conversely, as we show in Appendix \ref{Acone},
any manifold admitting such a vector field is a cone, at least locally.
Hence, the  conformal supercharges $S_{0}$ and $S_{5}$ remain
unbroken in a cone background and, similarly to the case of the
ordinary supercharges $Q_{0}$ and $Q_{5}$, they give rise to a family of TQFTs,
as we shall demonstrate explicitly in section \ref{SecOff}.

Using the vector field (\ref{conforv}),
the conformal scalar supersymmetries are given by
\be
\ba{ll}
{}\left[S_{0},A^{+}_{\mu}\right]=2\chi_{\mu\nu}\vx^{\nu}\,,~~~~&
\left[S_{5},A^{+}_{\mu}\right]=2\zeta\vx_{\mu}\,,\\
{}&{}\\
{}\left[S_{0},A^{-}_{\mu}\right]=2\eta\vx_{\mu}\,,~~~~&
\left[S_{5},A^{-}_{\mu}\right]=-2\left(\star\,\chi\right)_{\mu\nu}\vx^{\nu}\,,\\
{}&{}\\
{}\left[S_{0},\varphi\right]=-2\omega_{\mu}\vx^{\mu}\,,~~~~&
\left[S_{5},\varphi\right]=-2\psi_{\mu}\vx^{\mu}\,,\\
{}&{}\\
{}\left[S_{0},\bvarphi\right]=0\,,~~~~&
\left[S_{5},\bvarphi\right]=0\,,\\
{}&{}\\
{}\left[S_{0},h\right]=
2i\vx^{\mu}\cD^{+}_{\mu}\eta+4i\eta+\left[\omega_{\mu}\vx^{\mu},\bvarphi\right]\,,
~~~~&\left[S_{5},h\right]=
-2i\vx^{\mu}\cD^{-}_{\mu}\zeta-4i\zeta-\left[\psi_{\mu}\vx^{\mu},\bvarphi\right]\,,\\
{}&{}\\
{}\left\{S_{0},\chi_{\mu\nu}\right\}=\displaystyle{\frac{1}{\sqrt{g}}
\epsilon_{\mu\nu}{}^{\,\kappa\lambda}\vx_{\kappa}\cD^{-}_{\lambda}\bvarphi\,,}
~~~~&\left\{S_{5},\chi_{\mu\nu}\right\}
=-\vx_{\mu}\cD^{+}_{\nu}\bvarphi+\vx_{\nu}\cD^{+}_{\mu}\bvarphi\,,\\
{}&{}\\
\multicolumn{2}{l}{\left\{S_{0},\psi_{\mu}\right\}=\vx^{\lambda}\cH_{\lambda\mu}\!
+i\vx_{\mu}h+i\half\vx_{\mu}\left[\varphi,\bvarphi\right]-2i\phi_{\mu}\,,~
\left\{S_{5},\psi_{\mu}\right\}=-\left(\star\,F^{+}\right)_{\mu\nu}\vx^{\nu}\,,}\\
{}&{}\\
\left\{S_{0},\omega_{\mu}\right\}=-\left(\star\,F^{-}\right)_{\mu\nu}\vx^{\nu}\,,
  ~~~~&
\left\{S_{5},\omega_{\mu}\right\}=-\cH_{\mu\nu}\vx^{\nu}\!-i\vx_{\mu}h+
i\half\vx_{\mu}\left[\varphi,\bvarphi\right]+2i\phi_{\mu}\,,\\
{}&{}\\
{}\left\{S_{0},\eta\right\}=0\,,~~~~&
\left\{S_{5},\eta\right\}=\vx^{\mu}\cD_{\mu}^{-}\bvarphi+2\bvarphi\,,\\
{}&{}\\
{}\left\{S_{0},\zeta\right\}=\vx^{\mu}\cD_{\mu}^{+}\bvarphi+2\bvarphi\,,~~~~&
\left\{S_{5},\zeta\right\}=0\,,\\
{}&{}\\
{}\left[S_{0},g_{\mu\nu}\right]=0\,,~~~~&
\left[S_{5},g_{\mu\nu}\right]=0\,.
\ea
\label{ssalgebra}
\ee
\vfill\break
It is worth  noting that there exists a discrete  symmetry $\cZ$ given
by\footnote{
Note the similarity with the `Hermitian-conjugation' defined in \cite{Marcus:1995mq}.}
\be
\ba{lll}
\left(\ba{c}
A^{+}_{\mu}\\
A^{-}_{\mu}\\
\varphi\\
\bvarphi\\
h\\
\chi_{\mu\nu}\\
\psi_{\mu}\\
\omega_{\mu}\\
\eta\\
\zeta
\ea\right)~~&~~\stackrel{\cZ}{-\!\!\!-\!\!\!-\!\!\!\longrightarrow}~~&~~
\left(\ba{c}
A^{-}_{\mu}\\
A^{+}_{\mu}\\
\varphi\\
\bvarphi\\
-h\\
-\left(\star\,\chi\right)_{\mu\nu}\\
\omega_{\mu}\\
\psi_{\mu}\\
\zeta\\
\eta
\ea\right)\,,
\ea
\label{discrete}
\ee
under which the Lagrangian (\ref{topaction}) is invariant.
Furthermore $\cZ$ acts on the scalar supercharges as
\be
\ba{lll}
Q_{0}={\cZ\circ Q_{5}\circ\cZ}\,,~~~~&~~~~S_{0}={\cZ\circ S_{5}\circ\cZ}\,,
~~~~&~~~~\cZ^{2}=\mbox{identity}\,.
\ea
\ee
Apart from the four scalar supersymmetries, the discrete symmetry $\cZ$
and the diffeomorphisms of the three-dimensional base,
there exist another two  (bosonic) symmetries of
the twisted $\cN=4$ super Yang-Mills in a generic cone background.
One is the $\mbox{U}(1)$ $R$-symmetry or the
$\SO(2)$ rotation on the $\left(\Phi_{9},\Phi_{10}\right)$ plane
\be
\ba{ll}
{}\left[\Rs,A_{\mu}^{+}\right]=0\,,
~~~~~&~~~~~
{}\left[\Rs,A_{\mu}^{-}\right]=0\,,\\
{}&{}\\
{}\left[\Rs,\varphi\right]=+i\varphi\,,
~~~~~&~~~~~
{}\left[\Rs,\bvarphi\right]=-i\bvarphi\,,\\
{}&{}\\
{}\left[\Rs,h\right]=0\,,~~~~~&~~~~~
{}\left[\Rs,\chi_{\mu\nu}\right]=-i\half\chi_{\mu\nu}\,,\\
{}&{}\\
{}\left[\Rs,\psi_{\mu}\right]=+i\half\psi_{\mu}\,,
~~~~~&~~~~~
\left[\Rs,\omega_{\mu}\right]=+i\half\omega_{\mu}\,,\\
{}&{}\\
{}\left[\Rs,\eta\right]=-i\half\eta\,,~~~~~&~~~~~
{}\left[\Rs,\zeta\right]=-i\half\zeta\,,\\
{}&{}\\
{}\left[\Rs,g_{\mu\nu}\right]=0\,.~~~~&~~~~{}
\ea
\ee
\vfill\break
The other  is a `dilatation' symmetry\footnote{
Considering the asymmetric scaling dimensions of $\varphi$ and $\bvarphi$
in (\ref{dilatation}), one may wish to regard  $\fD+\Rs$ as the dilation operator instead.}
\be
\ba{ll}
{}\left[\fD,A_{\mu}^{+}\right]=v^{\lambda}\nabla_{\lambda}A^{+}_{\mu}+A^{+}_{\mu}\,,
~~~~&~~~~
{}\left[\fD,A_{\mu}^{-}\right]=v^{\lambda}\nabla_{\lambda}A^{-}_{\mu}+A^{-}_{\mu}\,,\\
{}&{}\\
{}\left[\fD,\varphi\right]=v^{\mu}\nabla_{\mu}\varphi\,,
~~~~&~~~~
{}\left[\fD,\bvarphi\right]=v^{\mu}\nabla_{\mu}\bvarphi+2\bvarphi\,,\\
{}&{}\\
{}\left[\fD,h\right]=v^{\mu}\nabla_{\mu}h+2h\,,~~~~&~~~~
{}\left[\fD,\chi_{\mu\nu}\right]=v^{\lambda}\nabla_{\lambda}\chi_{\mu\nu}
+2\chi_{\mu\nu}\,,\\
{}&{}\\
{}\left[\fD,\psi_{\mu}\right]=v^{\lambda}\nabla_{\lambda}\psi_{\mu}+\psi_{\mu}\,,
~~~~&~~~~
\left[\fD,\omega_{\mu}\right]=v^{\lambda}\nabla_{\lambda}\omega_{\mu}+\omega_{\mu}\,,\\
{}&{}\\
{}\left[\fD,\eta\right]=v^{\mu}\nabla_{\mu}\eta+2\eta\,,~~~~&~~~~
{}\left[\fD,\zeta\right]=v^{\mu}\nabla_{\mu}\zeta+2\zeta\,,\\
{}&{}\\
{}\left[\fD,g_{\mu\nu}\right]=0\,.~~~~&~~~~{}
\ea
\label{dilatation}
\ee
Since in flat space the inversion map (\ref{inversion}) transforms  the conformal supersymmetries to
the ordinary supersymmetries
and {vice versa} (\ref{inversionSQ}), one might wonder whether
the theory defined by the conformal supercharges $S_0$, $S_5$ is truly
inequivalent to the one given by the ordinary supercharges $Q_0$, $Q_5$.
In a flat background, a  straightforward computation shows that
the conformal  {scalar} supercharges  are related by the inversion map to the ordinary {vector}
supercharges. Schematically\footnote{More precisely:
\be
\ba{lll}
\left[S_{0}-{x^{\mu}Q_{\mu}}, h\right]=4i\eta\,,~~~&~~~
\left[S_{0}-{x^{\mu}Q_{\mu}}, \psi_{\nu}\right]=-2i\phi_{\nu}\,,~~~&~~~
\left[S_{0}-{x^{\mu}Q_{\mu}}, \zeta\right]=2\bvarphi\,,\\
{}[S_{5}-{x^{\mu}\tilde{Q}_{\mu}}, h]=-4i\zeta\,,  ~~~&~~~\,
{}[S_{5}-{x^{\mu}\tilde{Q}_{\mu}}, \omega_{\nu}]=2i\phi_{\nu}\,,~~~&~~~\,
{}[S_{5}-{x^{\mu}\tilde{Q}_{\mu}}, \eta]=2\bvarphi\,,
\ea\ee
with all other commutators identically zero.},
\be
\ba{ll}
S_{0}~\sim~{x^{\mu}Q_{\mu}} \,,~~~~~&~~~~~S_{5}~\sim~{x^{\mu}\tilde{Q}_{\mu}}\,.
\ea
\ee
Hence the conformal scalar supercharges are not equivalent to the
ordinary scalar supercharges.

\subsection{Superconformal algebra}
All scalar supercharges are  nilpotent off-shell
\be
\ba{llll}
Q_{0}^{2}=0\,,~~~~&~~~~Q_{5}^{2}=0\,,~~~~&~~~~S_{0}^{2}=0\,,~~~~&~~~~S_{5}^{2}=0\,.
\ea
\label{nilp}
\ee
Moreover, among the scalar supercharges there are two pairs which anticommute off-shell:
\be
\ba{ll}
\displaystyle{\left\{Q_{0},S_{5}\right\}=0\,,}~~~~&~~~~
\displaystyle{\left\{Q_{5},S_{0}\right\}=0\,.}
\ea
\label{anticom}
\ee
All other pairs of supercharges only anticommute on-shell.
More specifically, up to the equations of motion (\ref{EOM}) we have
\be
\ba{ll}
\displaystyle{\left\{Q_{0},Q_{5}\right\}\equiv -2\mathfrak{L}_{\varphi}}\,,~~~~&~~~~
\displaystyle{\left\{S_{0},S_{5}\right\}\equiv 2\mathfrak{L}_{v^{2}\bvarphi}}\,,\\
{}&{}\\
\displaystyle{\left\{Q_{0},S_{0}\right\}\equiv
-2\left(\fD-\mathfrak{L}_{v{\cdot A^{+}}}\right)}\,,~~~~&~~~~
\displaystyle{\left\{Q_{5},S_{5}\right\}\equiv
-2\left(\fD-\mathfrak{L}_{v{\cdot A^{-}}}\right)}\,,
\ea
\label{Allother}
\ee
where  $\mathfrak{L}_{v{\cdot A^{\pm}}}$   denotes
infinitesimal gauge transformations generated by  $v^{\mu}A_{\mu}^{\pm}$,
while $\fD$ corresponds to the dilatation~(\ref{dilatation}).
In the present paper `\,$\equiv$\,' denotes an on-shell equality, \textit{i.e.} an
equality up to equations of motion.
The dilation operator commutes with all the scalar supercharges off-shell
\be
\ba{llll}
{}\left[\fD,Q_{0}\right]=0\,,~~~&~~~
{}\left[\fD,Q_{5}\right]=0\,,~~~&~~~
{}\left[\fD,S_{0}\right]=0\,,~~~&~~~
{}\left[\fD,S_{5}\right]=0\,.
\ea
\ee
The $\mathrm{U}(1)$ charges of the supercharges can be read off of the following
commutation relations
\be
\ba{llll}
{}\left[\Rs,Q_{0}\right]=-iQ_{0}\,,~~~&~~~
{}\left[\Rs,Q_{5}\right]=-iQ_{5}\,,~~~&~~~
{}\left[\Rs,S_{0}\right]=-i\half S_{0}\,,~~~&~~~
{}\left[\Rs,S_{5}\right]=-i\half S_{5}\,.
\ea
\ee
Finally,  the $R$-symmetry commutes with the dilatation off-shell
\be
{}\left[\Rs,\fD\right]=0\,.
\ee
{}From the superconformal algebra above, we conclude
that the most general linear combination of the four scalar supercharges
which is nilpotent, is parameterized by four complex numbers subject to one constraint:
\be
\ba{ll}
\displaystyle{
w_{1}Q_{0}+w_{2}Q_{5}+w_{3}S_{0}+w_{4}S_{5}\,,}~~~~&~~~~w_{1}w_{3}+w_{2}w_{4}=0\,.
\ea
\ee
In particular, there are four possible pairs   of supercharges which are
nilpotent:
\be
\ba{llll}
z_1Q_{0}+ z_2Q_{5}\,,~~~~&~~~~z_1S_{0}+z_2S_{5}\,,~~~~&~~~~
z_1S_{0}+z_2Q_{5}\,,~~~~&~~~~
z_1S_{5}+z_2Q_{0}\,.
\label{lcomb}
\ea
\ee
In the above, $z_{1,2}$ are arbitrary complex numbers. As in \cite{Kapustin:2006pk}
we note that an overall constant is unimportant, hence each pair
of supercharges corresponds to a family of topological theories parameterized by
$\mathbb{CP}^1$.
For simplicity,  henceforth we  work in the patch $z_1\neq0$.\footnote{
The parameter $z:=z_2/z_1$ is related to
the parameter $t$ of \cite{Kapustin:2006pk} through
$it={(1-z)/(1+z)}$. In particular
$\left|z\right|=1$ corresponds to $t$ being real.}
The first linear combination in (\ref{lcomb}) was recently  investigated by
Kapustin and Witten~\cite{Kapustin:2006pk} and was shown to be relevant to  the
Langlands program. The second linear combination is the main focus of
the present paper.  For completeness, in Appendix \ref{ApBPS}
we give all the supersymmetric
conditions, \textit{i.e.} the BPS equations, arising from each linear combination
in (\ref{lcomb}).


\subsection{Vanishing theorems - weak version\label{subVanishingw}}
For the rest of this section
we shall focus on the conformal supersymmetry
\be
S_z:=S_{0}+zS_{5}~.
\label{szdef}
\ee
We shall
derive  certain important `vanishing theorems', which follow from a single algebraic identity.
Let us define
\be
\ba{l}
\displaystyle{\cV_{\mu\nu}(z):=\frac{1}{\sqrt{g}}
\epsilon_{\mu\nu}{}^{\,\kappa\lambda}\vx_{\kappa}\cD^{-}_{\lambda}\bvarphi-z\left(
\vx_{\mu}\cD^{+}_{\nu}\bvarphi-\vx_{\nu}\cD^{+}_{\mu}\bvarphi\right)\,,}\\
{}\\
\displaystyle{\cV_{\mu}(z):=\vx^{\lambda}\cH_{\lambda\mu}\!
+i\vx_{\mu}\cD_{\lambda}\phi^{\lambda}
+i\half\vx_{\mu}\left[\varphi,\bvarphi\right]-2i\phi_{\mu}-z
\left(\star\,F^{+}\right)_{\mu\nu}\vx^{\nu}\,,}\\
{}\\
\displaystyle{\hat{\cV}_{\mu}(z):=
-\left(\star\,F^{-}\right)_{\mu\nu}\vx^{\nu}
-z\left(\cH_{\mu\nu}\vx^{\nu}+i\vx_{\mu}\cD_{\lambda}\phi^{\lambda}-
i\half\vx_{\mu}\left[\varphi,\bvarphi\right]-2i\phi_{\mu}\right)\,,}\\
{}\\
\displaystyle{\cV_{\eta}(z):=z\left(v^{\mu}\cD_{\mu}^{-}\bvarphi+2\bvarphi\right)\,,}\\
{}\\
\displaystyle{\cV_{\zeta}:=\vx^{\mu}\cD_{\mu}^{+}\bvarphi+2\bvarphi\,.}
\ea
\label{cVq}
\ee
{}From the on-shell conformal supersymmetry transformations of
the fermions (\ref{ssalgebra}), we
see that an $S_z$-supersymmetric configuration must satisfy, by definition, the following conditions
(BPS equations)
\be
\cV_{\mu\nu}(z)=\cV_{\mu}(z)=\hat{\cV}_{\mu}(z)=\cV_{\eta}(z)=\cV_{\zeta}=0\,.
\ee
Or, in compact notation, $\left|\cV(z)\right\rangle=0$. Disregarding the boundary terms
we obtain the following algebraic identity involving
two arbitrary complex numbers $z,w$\,:
\be
\ba{l}
\langle\cV(w)|\cV(z)\rangle=\langle\cV(z)|\cV(w)\rangle^{\ast}\\
{}\\
\displaystyle{:=\int{\rm d^{4}}x\,\frac{\sqrt{g}}{\,2v^{2}}\,\tr\!\left[
\textstyle{\frac{1}{2}}\cV^{\mu\nu}(z)\cV_{\mu\nu}(w)^{\dagger}
+\cV^{\mu}(z)\cV_{\mu}(w)^{\dagger}
+\hat{\cV}{}^{\mu}(z)\hat{\cV}_{\mu}(w)^{\dagger}
+\cV_{\eta}(z)\cV_{\eta}(w)^{\dagger}
+\cV_{\zeta}\cV_{\zeta}^{\dagger}\right]}\\
{}\\
\displaystyle{\,\,=\int{\rm d^{4}}x\,
\left(1+z{w}^{\ast}\right)
\cL^{\mathrm{boson}}_{\mathrm{twisted}}
+\int{\rm d^{4}}x~\textstyle{\frac{1}{8}}\,\epsilon^{\kappa\lambda\mu\nu}
\tr\!\Big[zF^{+}_{\kappa\lambda}F^{+}_{\mu\nu}
+w^{\ast}F^{-}_{\kappa\lambda}F^{-}_{\mu\nu}\Big]\,,}
\ea
\label{vanishingid}
\ee
where $\cL^{\mathrm{boson}}_{\mathrm{twisted}}$
is the bosonic part of our twisted $\cN=4$  super Yang-Mills
Lagrangian (\ref{topaction}), after eliminating
all the auxiliary fields:
\be
\displaystyle{
\cL^{\mathrm{boson}}_{\mathrm{twisted}}=
\sqrt{g}\,\tr\!\left[
\textstyle{\frac{1}{4}}F^{+}_{\mu\nu}F^{-\mu\nu}
+\half\left(\cD_{\mu}\phi^{\mu}\right)^{2}+\textstyle{\frac{1}{4}}\cD^{+}_{\mu}\varphi
\cD^{-\mu}\bvarphi+\textstyle{\frac{1}{4}}\cD^{-}_{\mu}\varphi\cD^{+\mu}\bvarphi
+\textstyle{\frac{1}{8}}\left[\varphi,\bvarphi\right]^{2}\right]\,.}
\label{bLt}
\ee
Note that $\cL^{\mathrm{boson}}_{\mathrm{twisted}}$ is  positive semi-definite.
Eq.(\ref{vanishingid}) is an algebraic identity which holds for
  field configurations which  are smooth enough at the origin of the cone and
fall off fast enough at spatial infinity,
so that we can neglect the boundary terms.

For  $w=z$, the identity (\ref{vanishingid}) reduces to
\be
\displaystyle{\int_{\cone}
\cL^{\mathrm{boson}}_{\mathrm{twisted}}=
\frac{\langle\cV(z)|\cV(z)\rangle}{1+\left|z\right|^{2}}\,-\,\cT(z)\,,}
\label{vanishingid2}
\ee
where we have set
\be
\displaystyle{\cT(z):=
\frac{1}{\,8\left(1+\left|z\right|^{2}\right)}
\int{\rm d^{4}}x\,\epsilon^{\kappa\lambda\mu\nu}
\tr\!\Big[zF^{+}_{\kappa\lambda}F^{+}_{\mu\nu}
+z^{\ast}F^{-}_{\kappa\lambda}F^{-}_{\mu\nu}\Big]=
\mathrm{Re}\!\left[\displaystyle{\int_{\cone}}\frac{
\tr\!\left(zF^{+}\!\wedge F^{+}\right)}{
1+\left|z\right|^{2}}\right]\,.}
\ee
Note that $\cT(z)$ is real and topological, in that it does not
depend on the metric.
On the other hand, the choice $w^{\ast}=z^{-1}$ gives another
identity for the bosonic action:
\be
\displaystyle{\int_{\cone}
\cL^{\mathrm{boson}}_{\mathrm{twisted}}=\half
\left\langle\cV\left(z^{\ast-1}\right)|\cV(z)\right\rangle
-{\textstyle{\frac{1}{4}}}\int_{\cone}\tr\Big[
zF^{+}\!\wedge F^{+}+{z^{-1}}F^{-}\!\wedge F^{-}\Big]\,.}
\label{vanishingid3}
\ee
Combining (\ref{vanishingid2}) and (\ref{vanishingid3})
one  can eliminate the topological
quantities to obtain\footnote{Note that
$\langle\cV(z_{1})+\cV(z_{2})|\cV(w)\rangle:=\langle\cV(z_{1})|\cV(w)\rangle
+\langle\cV(z_{2})|\cV(w)\rangle$, etc.}
\be
\displaystyle{\left(\left|z\right|-\left|z\right|^{-1}\right)^{2}\int_{\cone}
\cL^{\mathrm{boson}}_{\mathrm{twisted}}=
\left\langle\cV(z)-\cV\left(z^{\ast-1}\right)|\cV(z)-
\cV\left(z^{\ast-1}\right)\right\rangle\,.}
\ee
{}From the positive semi-definite property of
$\cL^{\mathrm{boson}}_{\mathrm{twisted}}$ and
$\langle\cV(z)|\cV(z)\rangle$,  we have the following  inequalities
for any bosonic configuration:
\be
\displaystyle{\int_{\cone}
\cL^{\mathrm{boson}}_{\mathrm{twisted}}\,\geq\,
-\,\cT(z)\,\geq\,-\,
\frac{\langle\cV(z)|\cV(z)\rangle}{1+\left|z\right|^{2}}\,.}
\label{BPSbound}
\ee
Hence, given a complex number $z$,
a supersymmetric configuration (for which $\left|\cV(z)\right\rangle=0$)
minimizes the action within a given topological sector.
For {an arbitrary  bosonic configuration,}  regarding  $\cT(z)$
as a function of $z$, the minimum and the maximum of $\cT(z)$ are located
generically\footnote{Unless $\displaystyle{
\int_{\cone}\!\tr\!\left(F^{+}\!\wedge F^{+}\right)=0}$,
 in which case $\cT(z)$ is identically zero.}
at two antipodal points  on the unit circle:
\be
\ba{lll}
\cT(z_{\mathrm{min}})\leq\cT(z)\leq\cT(z_{\mathrm{max}})\,,~~~~&~~~~
\left|z_{\mathrm{min}}\right|=\left|z_{\mathrm{max}}\right|=1\,,~~~~&~~~~
z_{\mathrm{min}}+z_{\mathrm{max}}=0\,,
\ea
\ee
such that
\be
\displaystyle{
\cT(z_{\mathrm{min}})+\cT(z_{\mathrm{max}})=0\,.}
\ee
At the extremal  points,
$\displaystyle{\int_{\cone}\!{z}\,\tr\!\left(F^{+}\!\wedge F^{+}\right)}$
becomes real.
Hence
\be
\displaystyle{
\cT(z_{\mathrm{min}})
=\int_{\cone}\!\half{z_{\mathrm{min}}}
\tr\!\left(F^{+}\!\wedge F^{+}\right)=
\int_{\cone}\!\half{z_{\mathrm{min}}^{\ast}}
\tr\!\left(F^{-}\!\wedge F^{-}\right)\leq 0\,.}
\label{realcT}
\ee
Since  the left hand side of (\ref{vanishingid2})
is  independent of $z$, the above analysis implies that, for any given
bosonic configuration,
$\langle\cV(z)|\cV(z)\rangle/{(1+\left|z\right|^{2})}$
is bounded
both from  above and below:
\be
\ba{l}
\displaystyle{
\half\langle\cV(z_{\mathrm{min}})|\cV(z_{\mathrm{min}})\rangle
~\leq~\frac{\langle\cV(z)|\cV(z)\rangle}{1+\left|z\right|^{2}}
~\leq~\half\langle\cV(z_{\mathrm{max}})|\cV(z_{\mathrm{max}})\rangle\,,}\\
{}\\
\displaystyle{\langle\cV(z_{\mathrm{max}})|\cV(z_{\mathrm{max}})\rangle=
\langle\cV(z_{\mathrm{min}})|\cV(z_{\mathrm{min}})\rangle+4\cT(z_{\mathrm{max}})\,.}
\ea
\label{VVmin}
\ee
In particular, if $\langle\cV(z_{\mathrm{min}})|\cV(z_{\mathrm{min}})\rangle\neq 0$,
$\langle\cV(z)|\cV(z)\rangle$ cannot vanish for all $z$.
As in \cite{Kapustin:2006pk}, this leads to certain vanishing theorems.
\begin{itemize}
\item\textit{Vanishing theorem for generic configurations}:\label{pagevanishinggc}

A configuration is supersymmetric if and only if
$\langle\cV(z_{\mathrm{min}})|\cV(z_{\mathrm{min}})\rangle=0$.
\end{itemize}

\noindent {}Furthermore, from the identities (\ref{vanishingid}), (\ref{vanishingid2})
we obtain
the following vanishing theorems for supersymmetric configurations.
\begin{itemize}
\item\textit{Vanishing theorems for BPS configurations}:
\end{itemize}\begin{enumerate}
\item The choice $w^{\ast}=-z^{-1}$ in (\ref{vanishingid})
shows that any BPS state, for which $\left|\cV(z)\right\rangle=0$,
should satisfy
\be
\displaystyle{\int_{\mathrm{cone}} \tr\!\left(F^{-}\!\wedge F^{-}\right)=
\int_{\mathrm{cone}} z^{2}\, \tr\!\left(F^{+}\!\wedge F^{+}\right)\,.}
\ee
Hence, by complex conjugation,  we obtain
\be
\ba{ll}
\displaystyle{
\int_{\mathrm{cone}}  \tr\!\left(F^{+}\!\wedge F^{+}\right)=
\int_{\mathrm{cone}}  \tr\!\left(F^{-}\!\wedge F^{-}\right)=0\,,~~~~~\cT(z)=0}
~~~&~~~
\mathrm{if~~}\left|z\right|\neq 1\,,\\
{}&{}\\
\displaystyle{
\cT(z)=\int_{\mathrm{cone}}  \frac{z}{2}\,\tr\!\left(F^{+}\!\wedge F^{+}\right)=
\int_{\mathrm{cone}}  \frac{~z^{\ast}}{2}\,
\tr\!\left(F^{-}\!\wedge F^{-}\right)~~:~~\mathrm{real}}
~~~&~~~\mathrm{if~~}\left|z\right|= 1\,.
\ea
\label{vanishing1}
\ee
\item Conversely, when
$\displaystyle{\int  \tr\!\left(F^{+}\!\wedge F^{+}\right)\neq 0}$,
the BPS equations $\left|\cV(z)\right\rangle=0$
have no solution for  $\left|z\right|\neq 1$.
\item A BPS state, for which  $\left|\cV(z)\right\rangle=0$,
saturates the bound in (\ref{BPSbound})
\be
\displaystyle{\left.\int{\rm d^{4}}x\,
\cL^{\mathrm{boson}}_{\mathrm{twisted}}\right|_{\mathrm{BPS}}
=-\cT(z)=}\left\{
\ba{cl}
\displaystyle{
-z\!\int{\rm d^{4}}x\,{\textstyle{\frac{1}{8}}}\,
\epsilon^{\kappa\lambda\mu\nu}
\,\tr\!\left(F^{+}_{\kappa\lambda}F^{+}_{\mu\nu}\right)\,\geq\,0}
&~~~\mathrm{if~~}\left|z\right|=1\\
{}&{}\\
0&~~~\mathrm{otherwise\,.~}
\ea\right.
\ee
\item Any state which satisfies the BPS equations $\left|\cV(z)\right\rangle=0$
for some $z$ such that $\left|z\right|\neq 1$,
satisfies the BPS equations for all $z$.
\item Any state which satisfies the BPS equations $\left|\cV(z)\right\rangle=0$
for some $z$ such that  $\left|z\right|=1$,
satisfies the BPS equations for all $z$ iff $\cT(z)=0$.
%
\end{enumerate}
As a corollary we obtain an alternative criterion  for a generic configuration to be
supersymmetric.
\begin{itemize}
\item\textit{Corollary}:

A configuration is supersymmetric if and only if
$~|\cV(z)\rangle=0$ for some $\left|z\right|=1$.
\end{itemize}
Using explicitly the cone coordinates (\ref{cone}), the BPS equations read
for $\left|z\right|=1$:
\be
\ba{ll}
\displaystyle{
F^{+}_{jr}+z\left(\star\, F^{+}\right)_{jr}+2i\cD^{+}_{r}\phi_{j}+
i\frac{2}{r}\,\phi_{j}=0\,,}~~~~&~~~~
\displaystyle{\cD_{r}\varphi+\frac{2}{r}\varphi=0\,,}\\
{}&{}\\
\displaystyle{
\cD_{r}\phi_{r}+\frac{2}{r}\phi_{r}-\cD_{i}\phi^{i}=0\,,}~~~~&~~~~
\displaystyle{
\cD_{i}\varphi=0\,,}\\
{}&{}\\
\left[\phi_{\mu},\varphi\right]=0\,,~~~~&~~~~\left[\varphi,\bvarphi\right]=0\,.
\ea
\label{BPSz=1}
\ee
%
%
In particular if the BPS configuration has a trivial topology
$\cT(z)=0$ so that
$\displaystyle{\int\!\cL^{\mathrm{boson}}_{\mathrm{twisted}}=0}$,
it follows from (\ref{bLt}) that
the supersymmetric configuration satisfies
\be
\ba{lllll}
F^{+}_{\mu\nu}=0\,,~~~&~~~D_{\mu}\phi^{\mu}=0\,,~~~&~~~\varphi=0\,,~~~&~~~
\displaystyle{\cD_{r}\phi_{\mu}+\frac{1}{r}\phi_{\mu}=0\,,}~~~&~~~
\left[\phi_{r},\phi_{i}\right]=0\,.
\ea
\label{BPStrivial}
\ee
In the special case where  the base of the cone is the round sphere, the cone
 reduces to $\mathbb{R}^{4}$.  From (\ref{RicciA}) it then follows that
a BPS configuration with trivial topology  satisfies  $F_{\mu\nu}=0$, $\varphi=
\phi_{\mu}=0$.  These conditions are  more restrictive than the usual vacuum
equations $F_{\mu\nu}=0$, $\cD_{\mu}\Phi_{I}=0$, $\left[\Phi_{I},\Phi_{J}\right]=0$,
where $\Phi_{I}:=(\phi_{\mu},\,\varphi,\,\bvarphi)$.
This is due to our requirement that the fields should decay
fast enough at the boundary, and
 is consistent with the fact that the
conformal symmetry  remains unbroken if the VEVs of the scalars  vanish.
\subsection{Vanishing theorems - strong version\label{subVanishings}}
Taking equations (\ref{FFF}), (\ref{DDRF})
and the Bianchi identity for  the Riemann tensor  into account, we obtain
\be
\ba{l}
\displaystyle{
\epsilon^{\kappa\lambda\mu\nu}\tr\!\left(F^{+}_{\kappa\lambda}F^{+}_{\mu\nu}\right)
=\epsilon^{\kappa\lambda\mu\nu}\tr\!\left(F_{\kappa\lambda}F_{\mu\nu}\right)
-4\epsilon^{\kappa\lambda\mu\nu}\nabla_{\kappa}
\tr\Big(\phi_{\lambda}\cD_{\mu}\phi_{\nu}
-i\phi_{\lambda}F_{\mu\nu}+{\textstyle{\frac{2}{3}}\,}
\phi_{\lambda}\phi_{\mu}\phi_{\nu}\Big)\,,}\\
{}\\
\displaystyle{
\epsilon^{\kappa\lambda\mu\nu}\tr\!\left(F^{-}_{\kappa\lambda}F^{-}_{\mu\nu}\right)
=\epsilon^{\kappa\lambda\mu\nu}\tr\!\left(F_{\kappa\lambda}F_{\mu\nu}\right)
-4\epsilon^{\kappa\lambda\mu\nu}\nabla_{\kappa}
\tr\Big(\phi_{\lambda}\cD_{\mu}\phi_{\nu}
+i\phi_{\lambda}F_{\mu\nu}-{\textstyle{\frac{2}{3}}\,}
\phi_{\lambda}\phi_{\mu}\phi_{\nu}\Big)\,.}
\ea
\label{FpmF}
\ee
This implies that the topological term is real and it reduces to
the usual instanton number
\be
\displaystyle{
\int_{\,\cone}\!\tr\!\left(F^{+}\!\wedge F^{+}\right)=
\int_{\,\cone}\!\tr\!\left(F^{-}\!\wedge F^{-}\right)=
\int_{\,\cone}\!\tr\!\left(F\!\wedge F\right)\,\in\, 16\pi^{2}
\mathbb{Z}\,,}
\label{FwedgeFpm}
\ee
provided  we can neglect the boundary terms in (\ref{FpmF}).
This requirement is automatically satisfied in the case where
the VEVs of the scalars vanish at infinity.
The latter condition is
also sufficient  for the conformal symmetry to be unbroken.\footnote{Also in
 ordinary (untwisted)
super Yang-Mills  in flat space,
demanding  the action to be finite leads
to the condition that  $F_{\mu\nu}$, $D_{\mu}\Phi_{I}$, $\left[\Phi_{I},\Phi_{J}\right]$ should vanish
fast enough at infinity.
On the other hand, in the twisted theory
we are considering, the boundary conditions for a finite action are slightly different
from the ones in the untwisted case. Namely, what should vanish at infinity is
$F^{+}_{\mu\nu}$, $\cD_{\mu}\phi^{\mu}$, $\cD_\mu^{\pm}\varphi$ rather than
$F_{\mu\nu}$, $D_{\mu}\Phi_{I}$, $\left[\Phi_{I},\Phi_{J}\right]$.}

Here, we shall not discuss
the issue of the  boundary conditions any further;
instead we shall assume that (\ref{FwedgeFpm}) holds and examine the consequences.
In this case the minimum and maximum of $\cT(z)$
occur at $z=\pm 1$, so that
\be
\ba{lll}
\cT(z_{\mathrm{min}})\leq\cT(z)\leq\cT(z_{\mathrm{max}})\,,~~~~&~~~~
z_{\mathrm{min}}^{2}=z_{\mathrm{max}}^{2}=1\,,~~~~&~~~~
z_{\mathrm{min}}+z_{\mathrm{max}}=0\,,\\
{}&{}&{}\\
\multicolumn{3}{c}{\displaystyle{\cT(z_{\rmax})=-\cT(z_{\min})=
\left|\int_{\cone}\half\,\tr\!\left(F\!\wedge F\right)\right|\,.}}
\ea
\label{TTstr}
\ee
Consequently we have the following strong version of the vanishing theorems of the
previous subsections.
%
%
\begin{itemize}
\item\textit{Vanishing theorems - strong version}:
\end{itemize}\begin{enumerate}
\item
A configuration is supersymmetric if and only if
$\left|\cV(z)\rangle=0\right.$ for $z=+1$ or $z=-1$.
\item For a  generic supersymmetric configuration
the action  saturates the bound
\be
\displaystyle{\left.\int{\rm d^{4}}x\,
\cL^{\mathrm{boson}}_{\mathrm{twisted}}\right|_{\mathrm{BPS}}
=\left|\int_{\cone}\half\,\tr\!\left(F\!\wedge F\right)\right|\,\in\, 8\pi^{2}
\mathbb{N}\,.}
\label{satz=1}
\ee
\end{enumerate}
In particular, in the case where the cone is the flat
$\mathbb{R}^{4}$, we see from (\ref{RicciA}), (\ref{satz=1}) that
the BPS state should satisfy $D_{\mu}\Phi_{I}=0$,
$\left[\Phi_{I},\Phi_{J}\right]=0$. Thus, the BPS equations
(\ref{BPSz=1}) reduce to the instanton equations
$F_{\mu\nu}\pm\left(\star F\right)_{\mu\nu}=0$,
$\phi_{\mu}=\varphi=0$.
\vfill\break

\section{Off-shell formulation\label{SecOff}}
The conformal supersymmetry transformations  (\ref{ssalgebra}) given
in the previous section have  two unsatisfactory features. Firstly,
the anti-commutator relation $\left\{S_{0},S_{5}\right\}\equiv 2\fL_{v^{2}\bvarphi}$
only holds on-shell. Secondly,  the transformations depend explicitly on the
metric on the base of the  cone, $\hat{g}_{ij}(y)$.
In this section, we shall construct an off-shell formalism for the superconformal symmetry
which is  manifestly independent of the metric on the base.
Specifically,  we shall introduce additional auxiliary fields
and off-shell superconformal transformations generated by a new conformal supercharge $S'_{z}$,
such that the following relation holds:
\be
\ba{ll}
S'_{z}\equiv S_{z}=S_{0}+zS_{5}\,,
\ea
\ee
\textit{i.e.} the supercharge  $S'_{z}$ reduces on-shell to the supercharge  $S_{z}$
defined in (\ref{szdef}). In addition, $S'_{z}$ will be shown
to be nilpotent off-shell.
For simplicity of notation, in the following we suppress the prime
in $S'_{z}$.



\subsection{Off-shell algebra for conformal supersymmetries}
In addition to  $h$,
let us introduce the following new auxiliary fields
\be
\ba{lllllll}
h^{+}_{\mu\nu}\,,~~&~~h^{-}_{\mu\nu}\,,~~&~~
\bh^{+}_{\mu\nu}\,,~~&~~\bh^{-}_{\mu\nu}\,,~~&~~
h^{+}_{\mu}\,,~~&~~h^{-}_{\mu}~&~:~~\mbox{bosonic}\,,\\
{}&{}&{}&{}&{}&{}&{}\\
\multicolumn{6}{c}{\xi^{+}_{\mu\nu}\,,~~~~~~~~~~~~~
\xi^{-}_{\mu\nu}}~&~:~~\mbox{fermionic}\,,
\ea
\ee
where all tensor fields  are anti-symmetric
\be
\ba{lll}
h^{\pm}_{\mu\nu}=-h^{\pm}_{\nu\mu}\,,~~~~&~~~~
\bh^{\pm}_{\mu\nu}=-\bh^{\pm}_{\nu\mu}\,,~~~~&~~~~
\xi^{\pm}_{\mu\nu}=-\xi^{\pm}_{\nu\mu}\,.
\ea
\label{antisym}
\ee
We require  all the  auxiliary fields  to be orthogonal to the vector $v$
\be
\ba{llll}
h^{\pm}_{\mu\nu}\vx^{\nu}=0\,,~~~~&~~~~\bh^{\pm}_{\mu\nu}\vx^{\nu}=0\,,~~~~&~~~~
h^{\pm}_{\mu}\vx^{\mu}=0\,,~~~~&~~~~
\xi^{\pm}_{\mu\nu}\vx^{\nu}=0\,.
\ea
\label{orthogonal}
\ee
Furthermore,
the bosonic auxiliary fields are required to satisfy reality conditions similar to (\ref{reality})
\be
\ba{llll}
\left(h^{+}_{\mu\nu}\right)^{\dagger}=\bh^{-}_{\mu\nu}\,,~~~~&~~~~
\left(h^{-}_{\mu\nu}\right)^{\dagger}=\bh^{+}_{\mu\nu}\,,~~~~&~~~~
\left(h^{+}_{\mu}\right)^{\dagger}=h^{-}_{\mu}\,,~~~~&~~~~
\left(h\right)^{\dagger}=h\,.
\ea
\label{realityh}
\ee
We also introduce a second fermionic two form field $\hchi_{\mu\nu}=-\hchi_{\nu\mu}$.
Under the  discrete symmetry (\ref{discrete}) we have
\be
\ba{lllll}
h^{+}_{\mu\nu}\,\stackrel{\,\cZ}{\longleftrightarrow}\,h^{-}_{\mu\nu}\,,~~&~~
\bh^{+}_{\mu\nu}\,\stackrel{\,\cZ}{\longleftrightarrow}\,\bh^{-}_{\mu\nu}\,,~~&~~
h^{+}_{\mu}\,\stackrel{\,\cZ}{\longleftrightarrow}\,
h^{-}_{\mu}\,,~~&~~
\xi^{+}_{\mu\nu}\,\stackrel{\,\cZ}{\longleftrightarrow}\,\xi^{-}_{\mu\nu}\,,~~&~~
\chi_{\mu\nu}\,\stackrel{\,\cZ}{\longleftrightarrow}\,\hchi_{\mu\nu}\,.
\ea
\label{discreteA}
\ee
Let us also define the projection operator
\be
\ba{lll}
\displaystyle{
P^{\mu}{}_{\nu}:=\delta^{\mu}{}_{\nu}-\frac{v^{\mu}v_{\nu}}{v^{2}}\,,}~~~~&~~~~
P^{\mu}{}_{\nu}v^{\nu}=0\,,~~~~&~~~~
P^{\lambda}{}_{\mu}P^{\mu}{}_{\nu}=P^{\lambda}{}_{\nu}\,,
\ea
\ee
which projects onto subspaces orthogonal to the vector $v$.

\vfill\break

We are now ready to give the  off-shell transformations of all fields:
\be
\ba{l}
{}\left[S_{z},A^{+}_{\mu}\right]=2\chi_{\mu\nu}v^{\nu}+2z\zeta v_{\mu}\,,\\
{}\\
{}\left[S_{z},A^{-}_{\mu}\right]=2\eta v_{\mu}+2z\hchi_{\mu\nu}v^{\nu}\,,\\
{}\\
{}\left[S_{z},\varphi\right]=-2\left(\omega_{\mu}+z\psi_{\mu}\right)\vx^{\mu}\,,\\
{}\\
{}\left[S_{z},\bvarphi\right]=0\,,\\
{}\\
{}\left\{S_{z},\chi_{\mu\nu}\right\}=\bh^{-}_{\mu\nu}-z\left(
\vx_{\mu}\cD^{+}_{\nu}\bvarphi-\vx_{\nu}\cD^{+}_{\mu}\bvarphi\right)\,,\\
{}\\
{}\left\{S_{z},\hchi_{\mu\nu}\right\}=z\bh^{+}_{\mu\nu}-
\vx_{\mu}\cD^{-}_{\nu}\bvarphi+\vx_{\nu}\cD^{-}_{\mu}\bvarphi\,,\\
{}\\
\left\{S_{z},\psi_{\mu}\right\}=\vx^{\lambda}\cH_{\lambda\mu}\!
+i\vx_{\mu}h+i\half\vx_{\mu}\left[\varphi,\bvarphi\right]-2i\phi_{\mu}-zh^{+}_{\mu}\,,\\
{}\\
\left\{S_{z},\omega_{\mu}\right\}=-h^{-}_{\mu}+z\left(
-\cH_{\mu\nu}\vx^{\nu}\!-i\vx_{\mu}h+
i\half\vx_{\mu}\left[\varphi,\bvarphi\right]+2i\phi_{\mu}\right)\,,\\
{}\\
{}\left\{S_{z},\eta\right\}=z\left(\vx^{\mu}\cD_{\mu}^{-}\bvarphi+2\bvarphi\right)\,,\\
{}\\
{}\left\{S_{z},\zeta\right\}=\vx^{\mu}\cD_{\mu}^{+}\bvarphi+2\bvarphi\,,\\
{}\\
{}\left[S_{z},h\right]=
2i\vx^{\mu}\cD^{+}_{\mu}\eta+4i\eta+\left[\omega_{\mu}\vx^{\mu},\bvarphi\right]
-z\left(
2i\vx^{\mu}\cD^{-}_{\mu}\zeta+4i\zeta+
\left[\psi_{\mu}\vx^{\mu},\bvarphi\right]\right)\,,\\
{}\\
{\left[S_{z},h^{+}_{\mu\nu}\right]=\xi^{+}_{\mu\nu}\,,}\\
{}\\
{\left[S_{z},h^{-}_{\mu\nu}\right]=\xi^{-}_{\mu\nu}\,,}\\
{}\\
{\left[S_{z},\bh^{+}_{\mu\nu}\right]=-2i\left[
v^{2}\hchi_{\mu\nu}+v_{\mu}\hchi_{\nu\kappa}v^{\kappa}
-v_{\nu}\hchi_{\mu\kappa}v^{\kappa}\,,\,\bvarphi\right]\,,}\\
{}\\
{\left[S_{z},\bh^{-}_{\mu\nu}\right]=-2iz\left[
v^{2}\chi_{\mu\nu}+v_{\mu}\chi_{\nu\kappa}v^{\kappa}
-v_{\nu}\chi_{\mu\kappa}v^{\kappa}\,,\,\bvarphi\right]\,,}\\
{}\\
{\left[S_{z},h^{+}_{\mu}\right]=
2v^{\kappa}v^{\lambda}\cD^{+}_{\lambda}\hchi_{\mu\kappa}+4\hchi_{\mu\nu}v^{\nu}
-2v^{2}P_{\mu}{}^{\nu}
\left(\cD^{-}_{\nu}\zeta
-i\left[\psi_{\nu},\bar{\varphi}\right]\right)\,,}\\
{}\\
{\left[S_{z},h^{-}_{\mu}\right]=z
\Big(2v^{\kappa}v^{\lambda}\cD^{-}_{\lambda}\chi_{\mu\kappa}+4\chi_{\mu\nu}v^{\nu}
-2v^{2}P_{\mu}{}^{\nu}\left(\cD^{+}_{\nu}\eta-i\left[\omega_{\nu},\bvarphi\right]
\right)\Big)\,,}\\
{}\\
{\left\{S_{z},\xi^{+}_{\mu\nu}\right\}=
2iz\left[v^{2}\bvarphi\,,\,h^{+}_{\mu\nu}\right]\,,}\\
{}\\
{\left\{S_{z},\xi^{-}_{\mu\nu}\right\}=
2iz\left[v^{2}\bvarphi\,,\,h^{-}_{\mu\nu}\right]\,.}
\ea
\label{szalgebraA}
\ee
The nice feature of the above extended algebra is that
$S_{z}$ squares \textit{off-shell} to a gauge transformation:
\be
S_{z}^{2}=2z\fL_{v^{2}\bvarphi}\,.
\ee
Moreover, $S_{z}$ is manifestly independent of the metric on the base of the cone.

Under the discrete symmetry $\cZ$ (\ref{discrete}), (\ref{discreteA}),
$S_{z}$ transforms as
\be
\ba{lll}
S_{z}~~&~~\stackrel{\cZ}{-\!\!\!-\!\!\!-\!\!\!\longrightarrow}~~&~~
zS_{{1/z}}\,.
\ea
\ee
Note also that $S_{z}$  preserves the anti-symmetric and the  orthogonal
properties of the auxiliary fields, (\ref{antisym}) and (\ref{orthogonal}).

On-shell $\chi_{\mu\nu}$ and $\hchi_{\mu\nu}$ are related by
\be
\displaystyle{
-\hchi_{\mu\nu}\equiv
\left(\star\,\chi\right)_{\mu\nu}=\frac{1}{\,2\sqrt{g}\,}\,
\epsilon_{\mu\nu}{}^{\kappa\lambda}\chi_{\kappa\lambda}\,,}
\label{onchi}
\ee
and the auxiliary fields reduce to their on-shell values
\be
\ba{ll}
\displaystyle{
\left(h^{+}_{\mu\nu}\right)_{\on}
=\frac{1}{\sqrt{g}}\epsilon_{\mu\nu}{}^{\kappa\lambda}
v_{\kappa}\cD^{+}_{\lambda}\varphi\,,}~~~~~&~~~~~
\displaystyle{
\left(h^{-}_{\mu\nu}\right)_{\on}=
\frac{1}{\sqrt{g}}\epsilon_{\mu\nu}{}^{\kappa\lambda}
v_{\kappa}\cD^{-}_{\lambda}\varphi\,,}\\
{}&{}\\
\displaystyle{
\left(\bh^{+}_{\mu\nu}\right)_{\on}
=\frac{1}{\sqrt{g}}\epsilon_{\mu\nu}{}^{\kappa\lambda}
v_{\kappa}\cD^{+}_{\lambda}\bvarphi\,,}~~~~~&~~~~~
\displaystyle{
\left(\bh^{-}_{\mu\nu}\right)_{\on}
=\frac{1}{\sqrt{g}}\epsilon_{\mu\nu}{}^{\kappa\lambda}
v_{\kappa}\cD^{-}_{\lambda}\bvarphi\,,}\\
{}&{}\\
\displaystyle{
\left(h^{+}_{\mu}\right)_{\on}
=\left(\star\,F^{+}\right)_{\mu\nu}v^{\nu}\,,}~~~~~&~~~~~
\displaystyle{
\left(h^{-}_{\mu}\right)_{\on}
=\left(\star\,F^{-}\right)_{\mu\nu}v^{\nu}\,,}
\ea
\label{bosonicAux}
\ee
and
\be
\ba{l}
\displaystyle{
\left(\xi^{+}_{\mu\nu}\right)_{\on}
:=-\frac{2}{\sqrt{g}}\,\epsilon_{\mu\nu}{}^{\kappa\lambda}
v_{\kappa}\left(v^{\rho}\cD^{+}_{\lambda}\omega_{\rho}+\omega_{\lambda}
+zv^{\rho}\cD^{+}_{\lambda}\psi_{\rho}+z\psi_{\lambda}
-i\left[\left(\star\,\hchi\right)_{\lambda\rho}v^{\rho},\varphi\right]\right)\,,}\\
{}\\
\displaystyle{
\left(\xi^{-}_{\mu\nu}\right)_{\on}
:=-\frac{2}{\sqrt{g}}\,\epsilon_{\mu\nu}{}^{\kappa\lambda}
v_{\kappa}\left(v^{\rho}\cD^{-}_{\lambda}\omega_{\rho}+\omega_{\lambda}+
zv^{\rho}\cD^{-}_{\lambda}\psi_{\rho}+z\psi_{\lambda}
-iz\left[\left(\star\,\chi\right)_{\lambda\rho}v^{\rho},\varphi\right]\right)\,,}
\ea
\label{fermionicAux}
\ee
which are consistent with  (\ref{antisym}), (\ref{orthogonal}),
(\ref{realityh}), (\ref{discreteA}). Note that
there is some ambiguity in the expressions of the on-shell values of
the fermionic auxiliary fields $\xi^{\pm}_{\mu\nu}$ in (\ref{fermionicAux}),
due to the equations of motion for the fermions. However,
the precise expressions  in (\ref{fermionicAux}) will turn out to be  necessary
for the off-shell construction of the Lagrangian as we shall see below in (\ref{offLag}).

\vfill\break

\subsection{Off-shell Lagrangian }
Let us define
\be
\ba{ll}
V_{z}:=\displaystyle{\frac{\sqrt{g}}{\,2v^{2}\left(1-\left|z\right|^{2}\right)}\,}
&\tr\!\!
\left[-\textstyle{\frac{1}{4}}\chi^{\mu\nu}
\displaystyle{\left\{\left[S_{z},\chi_{\mu\nu}\right]
-\frac{1}{\sqrt{g}}\epsilon_{\mu\nu}{}^{\kappa\lambda}v_{\kappa}
\cD_{\lambda}^{-}\bvarphi
-z\left(\star\, \bh^{+}\right)_{\mu\nu}\right\}^{\dagger}}
-\eta\left[S_{z},\eta\right]^{\dagger}\right.\\
{}&{}\\
{}&
+\textstyle{\frac{1}{4}}\hchi^{\mu\nu}
\displaystyle{\left\{\left[S_{z},\hchi_{\mu\nu}\right]
-z\frac{1}{\sqrt{g}}\epsilon_{\mu\nu}{}^{\kappa\lambda}v_{\kappa}
\cD_{\lambda}^{+}\bvarphi-\left(\star\,\bh^{-}\right)_{\mu\nu}\right\}^{\dagger}}
+\zeta\left[S_{z},\zeta\right]^{\dagger}\\
{}&{}\\
{}&-\omega^{\mu}\left\{\left[S_{z},\omega_{\mu}\right]+
\left(1-\left|z\right|^{2}\right)\left(\star\, F^{-}\right)_{\mu\nu}v^{\nu}
+2izv_{\mu}\left(h-\cD_{\lambda}\phi^{\lambda}\right)\right\}^{\dagger}\\
{}&{}\\
{}&\left.\!\displaystyle{+\psi^{\mu}{\left\{\left[S_{z},\psi_{\mu}\right]
-\frac{1}{\,z^{\ast}}\left(1-\left|z\right|^{2}\right)
\left(\star\, F^{+}\right)_{\mu\nu}v^{\nu}
-2iv_{\mu}\left(h-\cD_{\lambda}\phi^{\lambda}\right)\right\}^{\dagger}}
}\right]\,.
\ea
\ee
Our main  formula is then
\be
-\textstyle{{\frac{1}{16}}}
\displaystyle{\,\epsilon^{\kappa\lambda\mu\nu}\,\tr\!\left[
z\,F^{+}_{\kappa\lambda}F^{+}_{\mu\nu}\,+\,\frac{1}{z}\,
F^{-}_{\kappa\lambda}F^{-}_{\mu\nu}\right]+
\Big\{S_{z}\,,\,V_{z}\Big\}
=\cL_{\mathrm{twisted}}\,+\,
\frac{\sqrt{g}}{\,v^{2}\left(1-\left|z\right|^{2}\right)}\,L_{\mathrm{auxiliary}}\,,}
\label{off-shellLag}
\ee
where $\cL_{\mathrm{twisted}}$ is the original Lagrangian given in (\ref{Stwisted})
and $L_{\mathrm{auxiliary}}$ consists of auxiliary terms
\be
\ba{ll}
&L_{\mathrm{auxiliary}}=\\
{}&{}\\
&~~~~\tr\displaystyle{\left[-\half\left(1-\left|z\right|^{2}\right)
\left\|h^{+}_{\mu}-\left(h^{+}_{\mu}\right)_{\on}\right\|^{2}
+\textstyle{\frac{1}{4}}\left|z\right|^{2}
\left\|h^{-}_{\mu\nu}-\left(h^{-}_{\mu\nu}\right)_{\on}\right\|^{2}
-\textstyle{\frac{1}{4}}
\left\|h^{+}_{\mu\nu}-\left(h^{+}_{\mu\nu}\right)_{\on}\right\|^{2}\right.}\\
{}&{}\\
{}&~~~~~~\displaystyle{~~+\textstyle{\frac{1}{8}}
\left(\chi+\star\,\hchi\right)^{\mu\nu}
\left(\xi^{+}_{\mu\nu}-\left(\xi^{+}_{\mu\nu}\right)_{\on}\right)
-\textstyle{\frac{1}{8}}z^{\ast}
\left(\hchi+\star\,\chi\right)^{\mu\nu}\left(
\xi^{-}_{\mu\nu}-\left(\xi^{-}_{\mu\nu}\right)_{\on}\right)}\\
{}&{}\\
{}&~~~~~~\left.\displaystyle{~~+\left(1-\left|z\right|^{2}\right)
\left(\hchi+\star\,\chi\right)^{\mu\rho}
{v_{\rho}\Lambda_{\mu\nu}v^{\nu}}}~\right]~+~\mathrm{total~derivatives}\,.
\ea
\label{offLag}
\ee
Here $\left(h^{+}_{\mu}\right)_{\on}$, $\left(h^{\pm}_{\mu\nu}\right)_{\on}$,
$\left(\xi^{\pm}_{\mu\nu}\right)_{\on}$ denote the on-shell
expressions of the auxiliary fields given in (\ref{bosonicAux}), (\ref{fermionicAux}),
and we have set
\be
\displaystyle{
\left\|T_{\mu_{1}\mu_{2}\cdots\mu_{n}}\right\|^{2}:=\frac{1}{\,n!}\,
T_{\mu_{1}\mu_{2}\cdots\mu_{n}}\left(T^{\mu_{1}\mu_{2}\cdots\mu_{n}}
\right)^{\dagger}\,.}
\label{normtensor}
\ee
We also define
\be
\displaystyle{
\Lambda_{\mu\nu}:=\cD^{+}_{\mu}\omega_{\nu}-\cD^{+}_{\nu}\omega_{\mu}
-\frac{1}{\sqrt{g}}\epsilon_{\mu\nu}{}^{\kappa\lambda}\cD^{-}_{\kappa}\psi_{\lambda}
+i\half\left[\left(\chi-\star\,\hchi\right)_{\mu\nu},\varphi\right]\,,}
\ee
which corresponds to the equation of motion for $\chi_{\mu\nu}$ (\ref{EOM}).
Thus all the terms in $L_{\mathrm{auxiliary}}$ vanish on-shell.

The equations of motion for the  auxiliary fields
$\xi^{\pm}_{\mu\nu}$, $h^{\pm}_{\mu}$, $h^{\pm}_{\mu\nu}$ give the
on-shell relations (\ref{onchi}) and (\ref{bosonicAux}). In particular,
a nice feature is that, due to the orthogonal property (\ref{orthogonal}),
we need to integrate out
both $\xi^{+}_{\mu\nu}$ and $\xi^{-}_{\mu\nu}$ in order to arrive at
the on-shell relation $\hchi_{\mu\nu}\equiv
-\left(\star\,\chi\right)_{\mu\nu}$ (\ref{onchi}). This is important  for a
nontrivial path integral measure for the fermionic auxiliary fields.
All other equations of motion then consist of those coming from
the original Lagrangian  (\ref{EOM}) together with the  additional equation
\be
\displaystyle{
\xi^{+}_{\mu\nu}-\left(\xi^{+}_{\mu\nu}\right)_{\on}=
\frac{z^{\ast}}{2\sqrt{g}}\,\epsilon_{\mu\nu}{}^{\kappa\lambda}
\Big[\xi^{-}_{\kappa\lambda}
\,-\,\left(\xi^{-}_{\kappa\lambda}\right)_{\on}\Big]\,,}
\ee
which is of course consistent with the on-shell relation (\ref{fermionicAux}).

{}From the above construction it follows
that the components of the energy-momentum tensor along the base are manifestly $S_z$-exact.
Hence the theory is topological, at least formally,
in the sense that it is independent of the metric on the base.

\section{Discussion\label{SecCoO}}
In this paper we focused on the twisting of four-dimensional $\cN=4$ SYM
studied in \cite{Marcus:1995mq,Blau:1996bx,
Labastida:1997vq,Labastida:1998sk,Lozano:1999ji, Kapustin:2006pk}.
We have found that putting the theory on a four-dimensional cone leaves
two conformal supercharges unbroken. The resulting theory is topological in that
it is independent of the metric on the base of the cone.

A few questions still remain
open. Firstly  one would like to identify the set of observables of the theory,
\textit{i.e.} the set of operators in the cohomology of the topological charge, and their
potential relevance to
topological invariants of the three-dimensional base of the cone.
The role of all  possible linear combinations of  nilpotent
scalar supercharges identified in section \ref{SecTw}, also deserves
further study.

The S-duality of $\cN=4$ SYM has significant implications for the
twisted version considered in the present paper, and this played a
central role in the recent discussion of \cite{Kapustin:2006pk}. It
would be interesting to examine this issue further. Another possible
direction would be to consider reductions of the theory to
lower-dimensional cones. A more thorough discussion of the different
possible boundary conditions is  postponed for future work.\newline
~\newline

\begin{center}
\large{\textbf{Acknowledgments}}
\end{center}
We are grateful to Giuseppe Policastro for useful discussions. JHP
benefited from  discussions with Nigel Hitchin and Nikita Nekrasov.
The research of JHP is supported by the Alexander von Humboldt
Foundation and  the Center for Quantum Spacetime  of Sogang
University with grant number R11 - 2005 - 021.
\newpage
\appendix
\section{Cone geometry\label{Acone}}
This section contains, among some other useful facts about cone geometry, a proof
of the statement that the existence of a vector field $v$ satisfying (\ref{conforv})
\be
\displaystyle{\nabla_{\mu}v^{\nu}=\delta_{\mu}{}^{\nu}\,,}
\label{conforv2}
\ee
implies that, at least locally, the manifold is a cone:
\be
\ba{lll}
{\rm{d}}s^{2}={\rm{d}}r^{2}+r^{2}{\rm{d}}\hat{s}^{2}\,,~~~~&~~~~
{\rm{d}}\hat{s}^{2}=\hat{g}_{ij}(y){\rm{d}}y^{i}{\rm{d}}y^{j}\,,~~~~&~~~~
y^{i}\neq r\,,
\ea
\ee
and the vector field is given by
\be
\displaystyle{v=r\frac{\,\partial~}{\,\partial r}\,.}
\label{vectorr}
\ee
~\\
\textit{Proof}\\
Let us define coordinates $x^{\mu}=\left(r,y^{i}\right)$ such that
\be
r:=\sqrt{v^{2\,}}\,.
\ee
Under diffeomorphisms of the form
\be
\ba{ll}
r~\longrightarrow~r^{\prime}=r\,,~~~~&~~~~
y^{i}~\longrightarrow~y^{\prime i}=f^{i}(r,y)\,,
\ea
\ee
the $g_{r i}(x)$ component  of the metric transforms as
\be
\displaystyle{g_{r i}(x)~\longrightarrow~g_{r i}^{\prime}(x)=
\frac{\partial f^{j}}{\partial y^{i}}\,g_{jk}(x^{\prime})\left(
\partial_{r}f^{k}(r,y)+\bar{g}^{kl}(x^{\prime})
g_{r l}(x^{\prime})\right)\,,}
\ee
where $\bar{g}^{kl}$ is the inverse of  $g_{ij}$,
$~\bar{g}^{ij}g_{jk}=\delta^{i}{}_{k}$.
Given initial conditions  $f^{k}(0,y)$ at  $r=0$,
one can uniquely fix the evolution along the $r$-direction by demanding
\be
\partial_{r}f^{k}(r,y)=-\bar{g}^{kl}\big(r,f(r,y)\big)
g_{r l}\big(r,f(r,y)\big)\,.
\ee
This  determines all the
higher-order derivatives of $f^{k}(r,y)$ with respect to $r$.
Thus, it is possible to choose a gauge such that~\cite{Kim:2006wg}
\be
g_{r i}=0\,.
\ee
Moreover, taking
 $\partial_{\mu}\left(r^{2}\right)=\partial_{\mu}\left(v^{2}\right)=2v_{\mu}$ into account,
we arrive at $v_{r}=r$, $v_{i}=0$. Hence, $g_{rr}=g^{rr}=1$ and the metric is of the form
\be
{\rm{d}}s^{2}={\rm{d}}r^{2}+r^{2}\hat{g}_{ij}\,{\rm{d}}y^{i}{\rm{d}}y^{j}\,.
\ee
Eq. (\ref{conforv2}) now reads
\be
\delta_{\mu}^{~\nu}=\nabla_{\mu}v^{\nu}=\delta_{\mu}^{~r}\delta_{r}^{~\nu}+
r\Gamma^{\nu}_{\mu r}\,.
\label{confor3v}
\ee
In particular, by taking the  $\nabla_{i}v^{j}$ component, we see that
$\hat{g}_{ij}$ is independent of the radial coordinate $r$.
Hence the geometry is a cone and the vector field is given by (\ref{vectorr}).
All other components of (\ref{confor3v}) can also be seen to hold, since:
\be
\ba{llll}
\displaystyle{
\Gamma^{i}{}_{rj}=\frac{1}{r}\delta^{i}_{~j}\,,}~&~
\Gamma^{r}_{ij}=-r\hat{g}_{ij}\,,~&~
\Gamma^{i}{}_{jk}=\half\hat{g}^{il}\left(\partial_{j}\hat{g}_{lk}
+\partial_{k}\hat{g}_{jl}-\partial_{l}\hat{g}_{jk}\right)\,,~&~
{\Gamma^{r}_{rr}=\Gamma^{r}_{ri}=\Gamma^{i}_{rr}}=0\,.
\ea
\label{chris}
\ee
This completes the proof.

Let us also mention the following useful relations:
\be
\ba{lll}
R_{rr}=0\,,~~~~&~~~~R_{ri}=R_{ir}=0\,,~~~~&~~~~R_{ij}=\hat{R}_{ij}-2\hat{g}_{ij}\,,
\ea
\ee
where $\hat{R}_{ij}$ is the Ricci curvature of the three-dimensional  base manifold.
In particular,
\be
v^{\mu}R_{\mu\nu}=0\,.
\ee

\section{BPS equations\label{ApBPS}}
{}It follows from the superconformal algebra of the twisted $\cN=4$ super Yang-Mills
(\ref{nilp}), (\ref{anticom}), (\ref{Allother}), that
there are four possible pairs of  supercharges which are
nilpotent, and hence define four different cohomological structures
\be
\ba{llll}
Q_{0}+ zQ_{5}\,,~~~~&~~~~S_{0}+z S_{5}\,,~~~~&~~~~S_{0}+zQ_{5}\,,~~~~&~~~~
S_{5}+zQ_{0}\,.
\ea
\ee
Here we give the BPS equations for each case, using
the cone coordinates (\ref{cone}). The equations follow directly from
(\ref{Qtransf}) and (\ref{ssalgebra}). Demanding that the fields vanish fast
enough at the boundary will generally result in additional constraints,  as in (\ref{BPStrivial}), but
we do not consider this issue here.
\subsection{$Q_{0}+ zQ_{5}$}
\begin{enumerate}
\item If $z=0$ :
\be
\ba{lll}
F^{+}_{\mu\nu}=0\,,~~~~&~~~~\cD^{+}_{\mu}\varphi=0\,,~~~~&~~~~
\cD_{\mu}\phi^{\mu}+\half\left[\varphi,\bvarphi\right]=0\,.
\ea
\label{z=0}
\ee
\item If $z=\infty$ :
\be
\ba{lll}
F^{+}_{\mu\nu}=0\,,
~~~~&~~~~\cD^{-}_{\mu}\varphi=0\,,~~~~&~~~~
\cD_{\mu}\phi^{\mu}-\half\left[\varphi,\bvarphi\right]=0\,.
\ea
\label{z=infty}
\ee
\item If $\left|z\right|=1$  :
\be
\ba{lllll}
F^{+}_{\mu\nu}-z\left(\star\, F^{-}\right)_{\mu\nu}=0
\,,~~&~~\cD^{}_{\mu}\varphi=0\,,~~&~~
\cD_{\mu}\phi^{\mu}=0\,,~~&~~\left[\phi_{\mu},\varphi\right]=0\,,
~~&~~\left[\varphi,\bvarphi\right]=0\,.
\ea
\ee
\item If $\left|z\right|\neq 0,\,\infty,\,1$  :
\be
\ba{lllll}
F^{+}_{\mu\nu}=0\,,~~~&~~~\cD^{}_{\mu}\varphi=0\,,~~~&~~~
\cD_{\mu}\phi^{\mu}=0\,,~~~&~~~\left[\phi_{\mu},\varphi\right]=0\,,
~~~&~~~\left[\varphi,\bvarphi\right]=0\,.
\ea
\ee
\end{enumerate}
\subsection{$S_{0}+z S_{5}$}
\begin{enumerate}
\item If $z=0$ :
\be
\ba{ll}
\displaystyle{
F^{+}_{jr}+2i\cD^{+}_{r}\phi_{j}+
i\frac{2}{r}\,\phi_{j}=0\,,}~~~~~&~~~~~F^{+}_{ij}=0\,,\\
{}&{}\\
\displaystyle{\cD^{-}_{r}\varphi+\frac{2}{r}\varphi=0\,,}~~~~~&~~~~~
\cD^{+}_{i}\varphi=0\,,\\
{}&{}\\
\displaystyle{
\cD_{r}\phi_{r}+\frac{2}{r}\phi_{r}-\cD_{i}\phi^{i}-\half
\left[\varphi,\bvarphi\right]=0\,.}~~~~~&~~~~~{}
\ea
\label{zz=0}
\ee
\item If $z=\infty$ :
\be
\ba{ll}
\displaystyle{
F^{+}_{jr}+2i\cD^{+}_{r}\phi_{j}+
i\frac{2}{r}\,\phi_{j}=0\,,}~~~~~&~~~~~F^{+}_{ij}=0\,,\\
{}&{}\\
\displaystyle{\cD^{+}_{r}\varphi+\frac{2}{r}\varphi=0\,,}~~~~~&~~~~~
\cD^{-}_{i}\varphi=0\,,\\
{}&{}\\
\displaystyle{\cD_{r}\phi_{r}+\frac{2}{r}\phi_{r}-\cD_{i}\phi^{i}+\half
\left[\varphi,\bvarphi\right]=0\,.}~~~~~&~~~~~{}
\ea
\label{zz=infty}
\ee
\item If $\left|z\right|=1$  :
\be
\ba{ll}
\displaystyle{
F^{+}_{jr}+z\left(\star\, F^{+}\right)_{jr}+2i\cD^{+}_{r}\phi_{j}+
i\frac{2}{r}\,\phi_{j}=0\,,}~~~~&~~~~
\displaystyle{\cD_{r}\varphi+\frac{2}{r}\varphi=0\,,}\\
{}&{}\\
\displaystyle{
\cD_{r}\phi_{r}+\frac{2}{r}\phi_{r}-\cD_{i}\phi^{i}=0\,,}~~~~&~~~~
\displaystyle{
\cD_{i}\varphi=0\,,}\\
{}&{}\\
\left[\phi_{\mu},\varphi\right]=0\,,~~~~&~~~~\left[\varphi,\bvarphi\right]=0\,.
\ea
\ee
\item If $\left|z\right|\neq 0,\,\infty,\,1$  :
\be
\ba{ll}
\displaystyle{
F^{+}_{jr}+2i\cD^{+}_{r}\phi_{j}+
i\frac{2}{r}\,\phi_{j}=0\,,}~~~~~&~~~~~F^{+}_{ij}=0\,,\\
{}&{}\\
\displaystyle{\cD_{r}\varphi+\frac{2}{r}\varphi=0\,,}~~~~~&~~~~~
\cD_{i}\varphi=0\,,\\
{}&{}\\
\displaystyle{\cD_{r}\phi_{r}+\frac{2}{r}\phi_{r}-\cD_{i}\phi^{i}=0
\,,}~~~~~&~~~~~
{\left[\varphi,\bvarphi\right]=0\,,}\\
{}&{}\\
{\left[\phi_{\mu},\varphi\right]=0\,.}~~~~~&~~~~~{}
\ea
\ee
\end{enumerate}
\subsection{$S_{0}+z Q_{5}$}
\be
\ba{ll}
r\cD^{+}_{j}\varphi+z^{\ast}F^{+}_{jr}=0\,,~~~~~~~~~~F^{+}_{ij}=0\,,
~~~~~&~~~~~
r\cD^{-}_{r}\varphi+2\varphi+iz^{\ast}\left(\cD_{\mu}\phi^{\mu}-\half
\left[\varphi,\bvarphi\right]\right)=0\,,\\
{}&{}\\
\displaystyle{
F^{+}_{jr}+2i\cD^{+}_{r}\phi_{j}+i\frac{2}{r}\phi_{j}-z\frac{1}{r}
\cD^{-}_{j}\varphi=0}\,,
~~~~&~~~~
\displaystyle{
\cD_{r}\phi_{r}+\frac{2}{r}\phi_{r}-\cD_{i}\phi^{i}
-\half\left[\varphi,\bvarphi\right]+
z\frac{i}{r}\cD^{-}_{r}\varphi=0\,.}
\ea
\label{zSQ}
\ee
If $z=0$ or $z=\infty$ the BPS equations reduce to (\ref{z=0}) or (\ref{z=infty}),
respectively.

\subsection{$S_{5}+z Q_{0}$}
\be
\ba{ll}
r\cD^{-}_{j}\varphi+z^{\ast}F^{-}_{jr}=0\,,~~~~~~~~~~F^{+}_{ij}=0\,,
~~~~~&~~~~~
r\cD^{+}_{r}\varphi+2\varphi-iz^{\ast}\left(\cD_{\mu}\phi^{\mu}+\half
\left[\varphi,\bvarphi\right]\right)=0\,,\\
{}&{}\\
\displaystyle{
F^{-}_{jr}-2i\cD^{-}_{r}\phi_{j}-i\frac{2}{r}\phi_{j}-z\frac{1}{r}
\cD^{+}_{j}\varphi=0}\,,
~~~~&~~~~
\displaystyle{\cD_{r}\phi_{r}+\frac{2}{r}\phi_{r}-\cD_{i}\phi^{i}
+\half\left[\varphi,\bvarphi\right]-
z\frac{i}{r}\cD^{+}_{r}\varphi=0\,.}
\ea
\label{zzSQ}
\ee
If $z=0$ or $z=\infty$ the BPS equations reduce to (\ref{z=infty}) or (\ref{z=0}),
respectively.

Note that under the discrete symmetry $\cZ$ (\ref{discrete}),
eqs. (\ref{z=0}),  (\ref{zz=0}), (\ref{zSQ})
transform to  eqs. (\ref{z=infty}), (\ref{zz=infty}),
(\ref{zzSQ}), respectively.
\section{Useful identities}
Any two-form tensors $A_{\mu\nu}$, $B_{\mu\nu}$ satisfy
\be
\displaystyle{A^{\mu\nu}v_{\nu}\left(\star\,B\right)_{\mu\lambda}v^{\lambda}+
\left(\star\,A\right)^{\mu\nu}v_{\nu}B_{\mu\lambda}v^{\lambda}=
\half v^{2}\left(\star\,A\right)^{\mu\nu}B_{\mu\nu}\,,}
\ee
which implies in particular
\be
\displaystyle{\left\{\chi^{\mu\nu}v_{\nu},
\left(\star\,\chi\right)_{\mu\lambda}v^{\lambda}\right\}=\textstyle{\frac{1}{4}}v^{2}
\left\{\chi^{\mu\nu},
\left(\star\,\chi\right)_{\mu\nu}\right\}\,.}
\ee
Some other useful identities are
\be
\displaystyle{
\frac{1}{\sqrt{g}}\epsilon_{\mu\nu}{}^{\kappa\lambda}
v_{\kappa}v^{\rho}\chi_{\lambda\rho}=-v^{2}\left(\star\,\chi\right)_{\mu\nu}
-v_{\mu}v^{\kappa}\left(\star\,\chi\right)_{\nu\kappa}
+v_{\nu}v^{\kappa}\left(\star\,\chi\right)_{\mu\kappa}\,,}
\ee
\be
\ba{l}
\displaystyle{\frac{1}{\,\left(v^{2}\right)^{2}}
\tr\!\left[v^{2}\phi_{\mu}v^{\mu}\cD_{\nu}\phi^{\nu}
+v^{2}\phi^{\mu}v^{\nu}\cD_{\mu}\phi_{\nu}+v^{2}\phi_{\mu}\phi^{\mu}
+2v^{\lambda}\phi_{\lambda}v^{\mu}v^{\nu}\cD_{\mu}\phi_{\nu}\right]}\\
{}\\
~~~~~~~~~~~~~~~~~~~~~~~~~~~~~~~~~~~~~~~~~~~~~~~~~~~~~~~~\displaystyle{=
\nabla_{\lambda}\tr\!\left[
\frac{\phi^{\lambda}\phi^{\mu}v_{\mu}}{v^{2}}+
\frac{v^{\lambda}\left(\phi^{\mu}v_{\mu}\right)^{2}}{\left(v^{2}\right)^{2}}\right]\,.}
\ea
\ee
For any two complex numbers, $z$, $w$, it follows from
(\ref{vanishingid2}), (\ref{off-shellLag}) that \be \displaystyle{
\int_{\cone}\Big\{S_{z}\,,\,V_{z}\Big\}=
\frac{\langle\cV(w)|\cV(w)\rangle}{1+\left|w\right|^{2}}+
\int_{\cone}\tr\Big(a_{+}\,F^{+}\!\wedge F^{+} +a_{-}\,F^{-}\!\wedge
F^{-}\Big)\,+\,\cdots\,,} \ee where we have set \be \ba{ll}
\displaystyle{a_{+}:=\frac{z}{4}-\frac{w}{\,2+2\left|w\right|^{2}}\,,}
~~~~&~~~~ \displaystyle{a_{-}:=
\frac{1}{\,4z\,}-\frac{w^{\ast}}{\,2+2\left|w\right|^{2}}\,,} \ea
\ee and the ellipses denote the fermionic and auxiliary parts; the
latter vanishes on-shell, as we can see from (\ref{onchi}),
(\ref{bosonicAux}), (\ref{fermionicAux}).
\vfill\break

\end{document}